\documentclass[twocolumn,floatfix]{revtex4-2}
\usepackage{graphicx} 
\usepackage{amssymb,epsfig,amsmath,amsfonts,bm,cases,physics,xcolor}
\usepackage{mathtools}
\usepackage{hyperref}
\mathtoolsset{showonlyrefs=false,showmanualtags=true}

\begin{document}
\title{Micromagnetic modelling and simulation of inertial magnetization dynamics in ferromagnets}
\author{Massimiliano d'Aquino}
\email{mdaquino@unina.it}
\affiliation{Department of Electrical Engineering and Information Technologies, University of Naples Federico II, 80125 Naples, Italy}











\begin{abstract}
Magnetization dynamics at sub-picosecond time scales can exhibit
inertial effects that lie beyond the conventional
Landau--Lifshitz--Gilbert description and give rise to an additional
nutational motion at terahertz frequencies. Following its theoretical
prediction and subsequent experimental observation in ferromagnets,
magnetic inertia has attracted increasing interest as a route toward
understanding and controlling ultrafast spin dynamics. In this topical
review, we discuss the theoretical modelling and micromagnetic
simulation of inertial magnetization dynamics in ferromagnetic
nanostructures. Starting from the mechanical analogy underlying the
inertial Landau--Lifshitz--Gilbert equation, we introduce its
spatially inhomogeneous micromagnetic formulation and discuss its main
qualitative properties, including magnetization-length conservation,
angular-momentum relaxation, and the generalized free-energy balance.
We then review the linear dynamics of inertial spin waves, highlighting
the coexistence of GHz precessional and THz nutational branches, the
inertial redshift of precessional modes, the spectral linewidths, and
the effects of exchange, magnetostatic interactions, confinement and
geometrical curvature and torsion. Particular attention is devoted to
the large-wavenumber regime, where magnetic inertia leads to a finite
limiting spin-wave propagation velocity. Finally, we address
frequency- and time-domain micromagnetic simulation techniques for
spatially nonuniform inertial dynamics, including geometric midpoint
integrators designed to preserve the fundamental constraints and
energy properties of the continuous dynamics. Representative
applications to the excitation and propagation of ultrashort
nutational spin waves and to ultrafast inertial magnetization switching
illustrate the role of micromagnetic modelling in connecting the
inertial Landau--Lifshitz--Gilbert theory with experimentally
accessible nanoscale magnetization dynamics.
\end{abstract}

\maketitle

\section{Introduction}

The investigation of magnetization dynamics on ultrashort
time scales has become one of the most active research
directions in modern magnetism, driven by the twofold
objective of understanding the fundamental physical
mechanisms governing spin dynamics far from equilibrium
and developing new generations of high-speed,
energy-efficient magnetic devices\cite{Chen2025}.
Access to magnetization processes occurring on
pico- and femtosecond time scales opens fascinating
possibilities for ultrafast information storage and
processing, including high-speed magnetic random-access
memories\cite{Bhatti2017}, unconventional and neuromorphic computing\cite{bracher2018,nakane2018,grollier_neuromorphic_2020},
and nanoscale sources and detectors of electromagnetic
signals in the terahertz (THz) frequency range\cite{Seifert2016THz,Chekhov2023THz}.
In this context, the development of ultrafast
spintronic\cite{Dieny2020opportunities} and magnonic devices\cite{kruglyak_magnonics_2010} may also offer
interesting opportunities for advanced communication
and signal-processing technologies\cite{chumak_roadmap_2022}, although their
practical realization still presents significant
scientific and technological challenges.

The modern field of ultrafast magnetism was largely
initiated by the pioneering experiments of Beaurepaire
et al.~\cite{Beaurepaire1996}, who demonstrated that
femtosecond optical excitation can induce demagnetization
of ferromagnetic nickel within approximately one
picosecond.
Subsequent time-resolved magneto-optical experiments
by Koopmans et al.~\cite{Koopmans2000} contributed
to clarifying the interplay between electronic,
optical and magnetic processes following ultrashort
laser excitation.
These investigations stimulated extensive research
on nonequilibrium spin dynamics and on the mechanisms
responsible for angular momentum transfer among
electronic, spin and lattice degrees of freedom.
Another important milestone was the demonstration
of inertia-driven spin switching in antiferromagnets
by Kimel et al.~\cite{Kimel2009}, showing that
ultrashort magnetic-field pulses can trigger
magnetization reorientation that continues after
the excitation has ceased.
These results provided compelling evidence that
magnetic dynamics at ultrashort time scales may
exhibit qualitatively different behavior from
the conventional precessional motion.

In parallel, the development of experimental
techniques capable of generating intense THz
electromagnetic pulses has opened new possibilities
for the direct excitation and observation of
coherent ultrafast magnetization dynamics.
In particular, the interaction of intense THz
magnetic fields with ferromagnetic thin films
has been investigated experimentally and
theoretically by Hudl et al.~\cite{Hudl2019},
demonstrating nonlinear magnetization dynamics
driven by strong THz excitation.
These developments are closely connected with
the rapidly growing field of ultrafast magnonics,
which aims at exciting, manipulating and detecting
high-frequency collective spin excitations
on nanometer spatial scales, as discussed
in the recent magnonics roadmap
by Flebus et al.~\cite{Flebus2024}.
In this perspective, understanding the propagation
and interaction of ultrashort-wavelength spin waves
at THz frequencies is becoming increasingly
important for both fundamental investigations
and prospective applications.

The theoretical description of magnetization dynamics
in ferromagnets has traditionally relied on the
Landau--Lifshitz--Gilbert (LLG) equation, which describes
the precessional motion of magnetization around the local
effective field together with dissipation through the
Gilbert damping mechanism.
The phenomenological formulation introduced by Gilbert
in 1955 and subsequently republished in
Ref.~\cite{gilbert_phenomenological_2004} was developed
by exploiting an analogy between magnetization dynamics
and the rotational motion of a rigid body.
An important, although often overlooked, aspect of
Gilbert's treatment concerns the assumption made on the
inertial tensor, for which the transverse components are
set to zero while the longitudinal component remains
finite. As discussed in detail in Sec.~II, this assumption
eliminates the contribution of magnetic inertia to the
equation of motion.

This assumption, for a single-domain nanomagnet
(macrospin) of volume $V$ with magnetic moment
$\bm M V=-\gamma_e \bm L$ fulfilling the semiclassical
relation between magnetization $\bm M$ and electron spin
angular momentum $\bm L$ through the gyromagnetic ratio
(here
$\gamma_e=g_e|q_e|/(2m_e)\approx176\times10^9\,
\mathrm{rad\,s^{-1}\,T^{-1}}$
is the absolute value, where $g_e\approx2$,
$|q_e|\approx1.6\times10^{-19}\,\mathrm{C}$ and
$m_e\approx9.1\times10^{-31}\,\mathrm{kg}$ are the
Land\'e factor, the magnitude of the electron charge and
the electron mass, respectively), leads to the classical
Landau--Lifshitz--Gilbert equation:
\begin{equation}
    \dv{\bm M}{t}
    =
    -\gamma \bm M\times
    \left[
        \bm H_\mathrm{eff}
        -\frac{\alpha}{\gamma M_s}
        \dv{\bm M}{t}
    \right] \,,
    \label{eq:LLG dimensional}
\end{equation}
where $\gamma=\mu_0\gamma_e$ is the gyromagnetic ratio
measured in $\mathrm{m\,A^{-1}\,s^{-1}}$, $M_s(T)$ is
the saturation magnetization of the ferromagnetic material
at temperature $T$, $\alpha$ is the dimensionless Gilbert
damping, and $\bm H_\mathrm{eff}$ is the magnetic effective
field accounting for the relevant magnetic interactions,
including magnetostatic, magnetocrystalline anisotropy and
Zeeman contributions.

\begin{figure*}[t]
    \centering
    \includegraphics[width=0.28\linewidth]{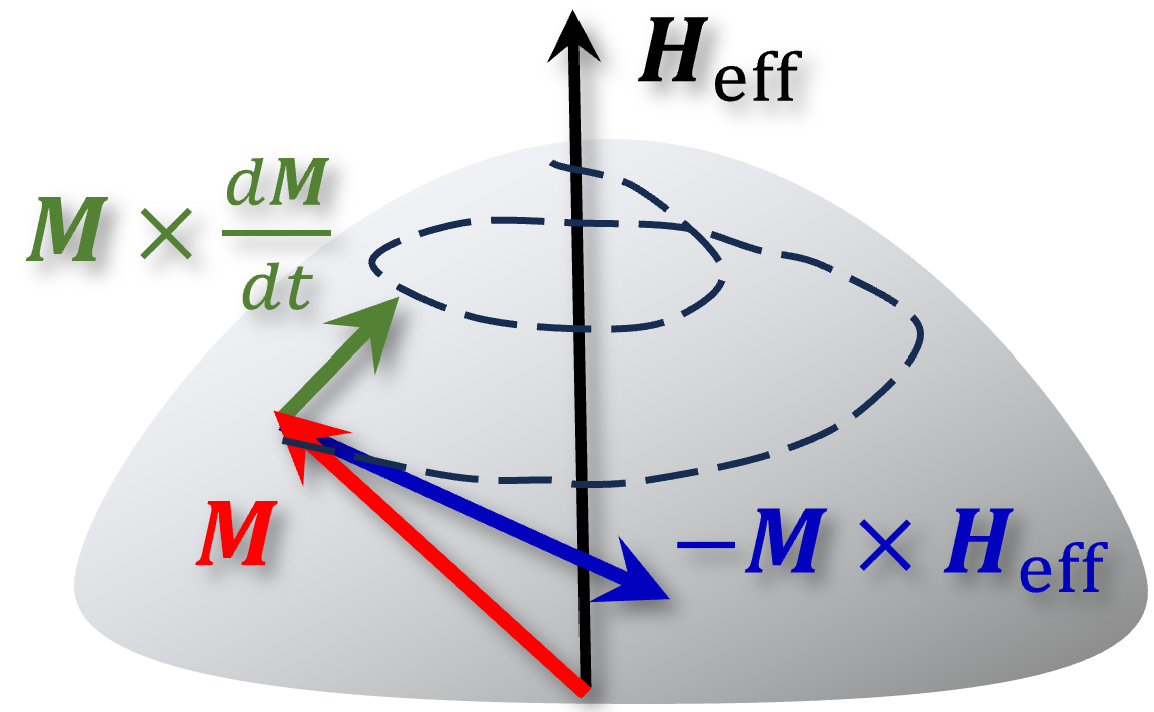}
    \hfill
    \includegraphics[width=0.34\linewidth]{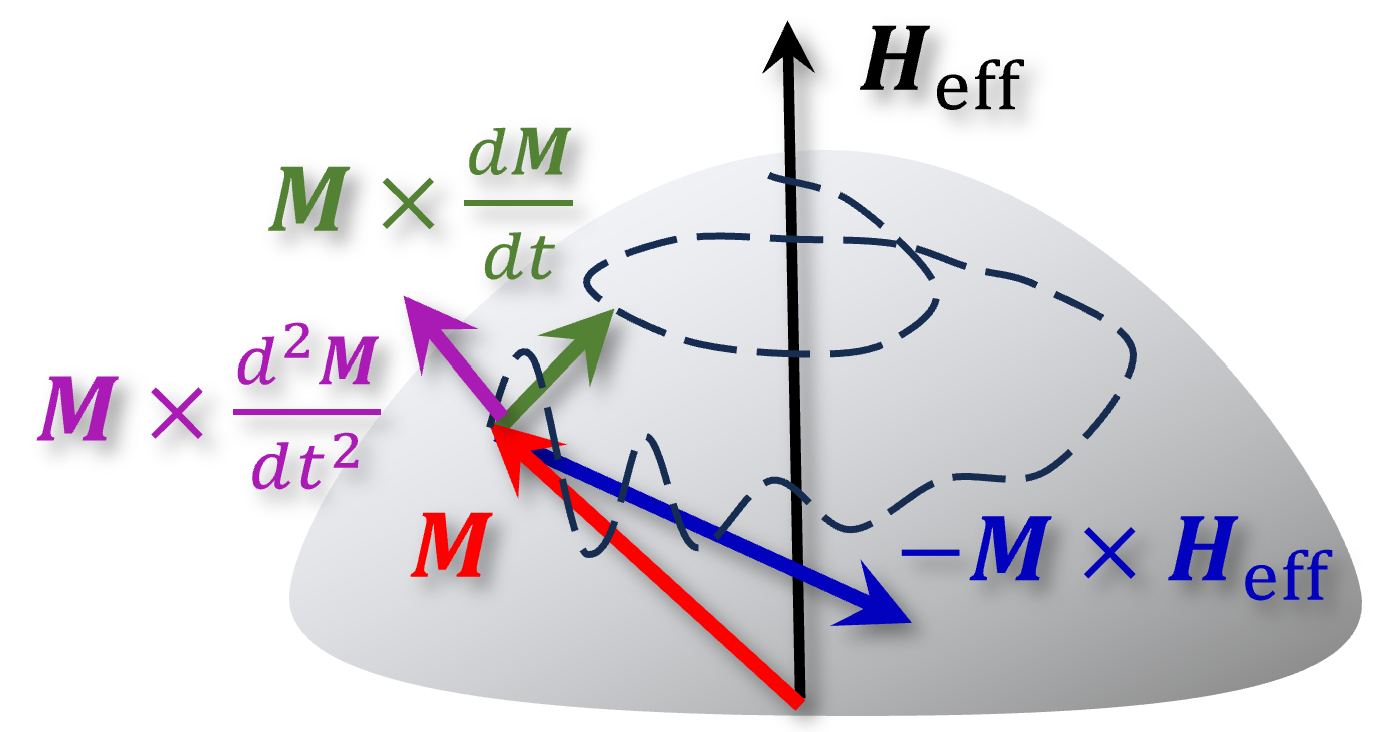}
    \hfill
    \includegraphics[width=0.22\linewidth]{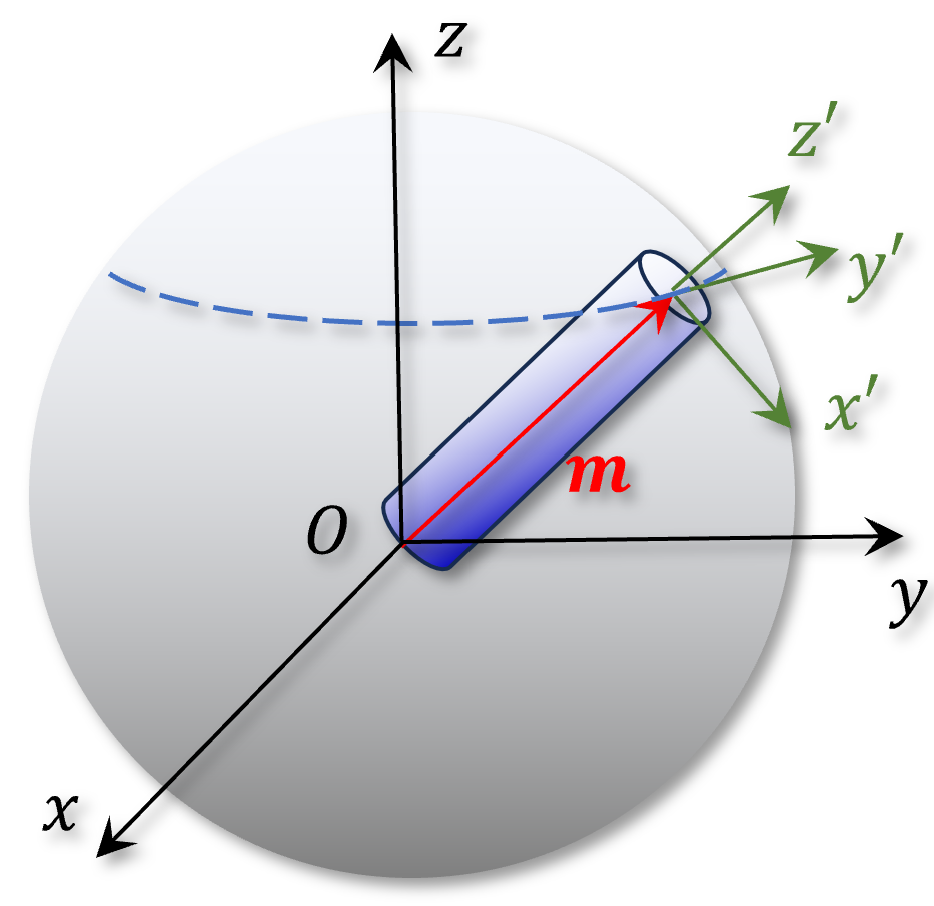}\\
    (a) \hspace{6cm}(b) \hspace{6cm} (c)
    \caption{Sketch of magnetization motion and effective
    magnetic torques in: (a) LLG equation
    \eqref{eq:LLG dimensional}; and (b) inertial LLG (iLLG)
    equation \eqref{eq:iLLG dimensional}. The torque terms
    $-\bm M\times\bm H_\mathrm{eff}$ (blue) drive
    magnetization precession around the effective field,
    damping terms $\bm M\times\dv{\bm M}{t}$ (green) push
    magnetization to align with the effective field towards
    equilibrium. For iLLG dynamics, the additional torque
    term $\bm M\times\dv[2]{\bm M}{t}$ (purple) drives
    ultrafast magnetization nutation at THz frequencies.
    (c) Sketch of a gyroscope (spinning rigid stick with one
    fixed end at the origin $O$) directed along the
    unit-vector $\bm m$. Lab frame axes are denoted as
    $x,y,z$ with unit vectors $\bm e_x,\bm e_y,\bm e_z$,
    while body-fixed frame axes are $x',y',z'$ with unit
    vectors $\bm e_x',\bm e_y',\bm e_z'$. The body-fixed
    frame is assumed to be rotating with angular velocity
    $\bm\Omega$.}
    \label{fig:magnetic torques}
\end{figure*}

The LLG equation~\eqref{eq:LLG dimensional} is the most
widespread theoretical tool used to analyze magnetization
dynamics. It is a torque equation acting on the
constant-amplitude magnetization,
$|\bm M(t)|=M_s$, and contains a conservative precessional
torque $-\bm M\times\bm H_\mathrm{eff}$, which drives
magnetization precession around the effective field, and a
dissipative Gilbert torque proportional to
$\bm M\times\dv{\bm M}{t}$, which drives the magnetization
toward equilibrium. A sketch of the typical magnetization
motion on the sphere of radius $M_s$ described by the LLG
equation is reported in Fig.~\ref{fig:magnetic torques}(a).

A fundamental theoretical development toward extending
this conventional picture was provided by Ciornei,
Rub\'i and Wegrowe~\cite{Ciornei2011magnetization}, who
reconsidered ferromagnetic magnetization dynamics by
enlarging the relevant phase space to include angular
momentum as an independent dynamical variable. Using
mesoscopic nonequilibrium thermodynamics, they derived a
generalized LLG equation containing an additional inertial
torque and predicted the emergence of magnetization
nutation at ultrashort time scales. The physical
interpretation of this formulation was further elucidated
by Wegrowe and Ciornei~\cite{wegrowe2012magnetization},
who discussed the connection between the gyromagnetic
relation, the mechanical analogy with a spinning rigid
body, and the occurrence of inertial effects in
magnetization dynamics.

Within this framework, the sole relaxation of Gilbert's
assumption on the inertial tensor produces an additional
angular-momentum relaxation term associated with magnetic
inertia, leading to the inertial Landau--Lifshitz--Gilbert
(iLLG) equation
\begin{equation}
    \dv{\bm M}{t}
    =
    -\gamma \bm M\times
    \left[
        \bm H_\mathrm{eff}
        -\frac{\alpha}{\gamma M_s}
        \dv{\bm M}{t}
        -t_\mathrm{in}^2
        \dv[2]{\bm M}{t}
    \right] \,,
    \label{eq:iLLG dimensional}
\end{equation}
where $t_\mathrm{in}$ is the characteristic time scale of
magnetic inertia, which typically amounts to fractions of
a picosecond. The additional torque proportional to
$\bm M\times\dv[2]{\bm M}{t}$ drives an ultrafast
nutational motion at THz frequencies superimposed on the
ordinary precessional dynamics occurring at much slower
GHz time scales, as schematically illustrated in
Fig.~\ref{fig:magnetic torques}(b). The mechanical origin
of this term and its connection with the gyroscope analogy
shown in Fig.~\ref{fig:magnetic torques}(c) will be
discussed in detail in Sec.~II.


The possibility of magnetic inertia was also investigated from different microscopic and theoretical perspectives. Fähnle et al.~\cite{fahnle2011generalized} derived an inertial extension of the Gilbert equation within an extended breathing Fermi surface model, establishing a connection between magnetic inertia and electronic relaxation mechanisms. Bhattacharjee et al.~\cite{bhattacharjee2012atomistic} developed a microscopic approach to atomistic spin dynamics incorporating exchange interactions, damping and magnetic inertia within a generalized equation of motion. Kikuchi and Tatara~\cite{kikuchi2015spin} derived spin inertia in metallic ferromagnets from a microscopic itinerant-electron model and investigated its consequences for spin-wave and domain-wall dynamics. Subsequently, Thonig et al.~\cite{thonig2017magnetic} investigated the magnetic moment of inertia within the torque--torque correlation formalism, providing an electronic-structure-based description of inertial effects and their relationship with Gilbert damping. In a different theoretical framework, Mondal et al.~\cite{Mondal2017} derived inertial contributions to magnetization dynamics from a relativistic quantum-mechanical description, connecting magnetic inertia with higher-order spin--orbit coupling effects.

A major experimental breakthrough occurred with the
observation of intrinsic inertial spin dynamics in
ferromagnetic thin films reported by Neeraj et
al.~\cite{Neeraj2021inertial}. Using intense THz
magnetic-field excitation, the authors demonstrated the
occurrence of magnetization nutation at frequencies in the
sub-THz range, providing direct experimental evidence of
the additional dynamical mode predicted by the inertial
extension of the LLG equation approximately one decade
earlier. This result renewed considerable interest in the
theoretical and experimental investigation of magnetic
inertia in ferromagnets. Further experimental evidence was
subsequently obtained in epitaxial cobalt films by
Unikandanunni et al.~\cite{Unikandanunni2022}, where
inertial spin dynamics was investigated in a system with
well-defined crystalline structure and magnetic
anisotropy.
More recently, De et al.~\cite{de2025magnetic}
investigated the transient separation between
magnetization and angular momentum following
ultrashort optical excitation of a permalloy
film, providing further experimental evidence
of inertial nutation at sub-THz frequencies.

In parallel with these experimental developments, further theoretical investigations have contributed
to clarifying the physical origin and broader
dynamical consequences of magnetic inertia.
Giordano and D\'ejardin~\cite{giordano2020derivation}
derived inertial magnetization dynamics from
the classical mechanics of a circular current loop,
providing an alternative mechanical interpretation
of the additional inertial torque.
More recently, Ghosh et
al.~\cite{ghosh2024theory} investigated the
generalization of magnetic inertia from a scalar
parameter to a tensorial quantity, highlighting
the possible role of anisotropic inertial effects.
The influence of magnetic inertia on nonlinear
magnetization dynamics has also been investigated,
including the emergence of nutational
auto-oscillations in ferromagnets
\cite{rodriguez2024spin}, nonlinear response
and dynamic magnetic hysteresis
\cite{titov2024nonlinear}, and inertia-induced
self-oscillations in antiferromagnetic systems
driven by spin-orbit torques
\cite{he2023large,he2024influence}.
Recent microscopic approaches have also
explored the possible role of coupling
between spin and lattice degrees of freedom
in generating effective magnetic inertia
\cite{webb2026microscopic}. Another recent microscopic interpretation was proposed
by Moussa et al.~\cite{moussa2026unquenched}, who
investigated the role of unquenched orbital angular
momentum as a possible origin of spin inertia.
Starting from coupled spin and orbital degrees of
freedom, they derived an effective inertial
magnetization equation and obtained an estimate of
the inertia parameter consistent with experimental
values reported for cobalt.

Considerable theoretical attention has also been devoted to the consequences of magnetic inertia for ferromagnetic resonance and spin-wave dynamics. In particular, the inertial LLG (iLLG) equation predicts the coexistence of conventional precessional modes and additional high-frequency nutational modes, which give rise to a new class of collective spin excitations, commonly referred to as nutational spin waves or inertial magnons. Their dispersion relations, linewidths and propagation properties have been investigated in several magnetic configurations, including extended films, confined nanostructures and anisotropic ferromagnets \cite{mondal2022inertial,cherkasskii2021dispersion,lomonosov2021anatomy,titov2022nutation}. Inertial effects have also been investigated in spin-wave dynamics associated with nonuniform magnetic textures, including magnetic vortices \cite{Gareeva2023}, as well as in the collective dynamics of skyrmions \cite{makhfudz2025controllable}. Of particular interest is the possibility of exciting and propagating inertial spin waves with nanometer wavelengths and THz frequencies, exhibiting distinctive features such as a finite limiting group velocity at large wavenumbers \cite{daquino2023micromagnetic} and nonreciprocal propagation in geometrically chiral nanostructures \cite{dAquino2025}.

The investigation of such phenomena requires
the extension of conventional micromagnetic
modelling\cite{brown_micromagnetics_1963,Abert2019} beyond the classical LLG equation.
Indeed, while the macrospin approximation
provides a useful description of spatially
uniform precession and nutation, the analysis
of inertial spin-wave excitation, propagation
and interference in realistic nanostructures
requires a spatially resolved treatment
including exchange, magnetostatic interactions,
magnetic anisotropy and external excitation.
Furthermore, the presence of the second-order
time derivative in the iLLG equation introduces
additional dynamical degrees of freedom and
poses specific challenges for numerical
simulation, particularly when resolving
THz oscillations and nanometer-scale
magnetization patterns.

In this context, frequency-domain micromagnetic
techniques provide efficient tools for
investigating linear inertial spin-wave
dynamics and resonance phenomena, while
time-domain simulations are essential for
studying transient propagation, nonlinear
magnetization dynamics and ultrafast switching.
Recent developments in geometric numerical
integration have also made it possible to
construct time-stepping methods that preserve
the fundamental constraints and energy
properties of the iLLG dynamics, providing
reliable computational tools for large-scale
inertial micromagnetic simulations
\cite{daquino2023micromagnetic,
daquino2024midpoint}.


The aim of this topical review is to present a unified overview of the theoretical modelling and numerical simulation of inertial magnetization dynamics in ferromagnets, with particular emphasis on spatially inhomogeneous magnetization processes and inertial spin waves. We first revisit the mechanical analogy underlying the introduction of magnetic inertia in Sec.~\ref{sec:brief review iLLG macrospin} and discuss the formulation and fundamental qualitative properties of the iLLG equation in Sec.~\ref{sec:inertial dynamics inhomogeneous}. We then review in Sec.~\ref{sec:modelling inertial SW dynamics} the existing literature and the analytical description of inertial spin-wave dynamics in macrospin systems, extended magnetic thin films and curved nanostructures, highlighting the emergence of nutational branches, their spectral properties and the limiting propagation velocity. Finally, in Sec.~\ref{sec:numerical}, we review the relevant numerical literature and discuss frequency- and time-domain micromagnetic simulation methods for the iLLG equation, with particular attention to geometric time integration and representative applications involving ultrashort-wavelength inertial spin waves and ultrafast magnetization switching.

\section{Brief review of inertial dynamics for a macrospin nanomagnet}\label{sec:brief review iLLG macrospin}

Here we review the essential concepts from the seminal
papers~\cite{Ciornei2011magnetization,wegrowe2012magnetization} needed to understand
the inertial dynamics of a macrospin (single-domain)
nanomagnet. To this end, we introduce the mechanical analogy between a uniformly-magnetized ferromagnet having magnetization $\bm M(t)=M_s\bm m(t)$ with constant length $|\bm M|=M_s$, directed along the unit-vector $\bm m(t)$ and a gyroscope, namely a spinning rigid stick with one fixed end at the origin $O$ as depicted in Fig.~\ref{fig:magnetic torques}(c).


One of the fundamental results of rigid body kinematics states that the velocities in the fixed and rotating reference frames obey the following composition rule:
\begin{equation}
    \dv{\bm M}{t}\bigg|_\mathrm{lab}=\dv{\bm M}{t}\bigg|_\mathrm{rot}+\bm\Omega\times \bm M \quad, \label{eq:velocity composition}
\end{equation}
where $\bm \Omega$ is the (instantaneous) angular velocity vector. For a rigid stick with constant length, it happens
\begin{equation}
    \dv{\bm M}{t}\bigg|_\mathrm{rot}=0 \quad.
\end{equation}
Then, by cross-multiplying both sides of Eq.~\eqref{eq:velocity composition}, one has:
\begin{align}
 \bm M \times  \dv{\bm M}{t} &= \bm M\times (\bm \Omega\times \bm M) = \\
 &= \bm \Omega(\bm M\cdot \bm M)-\bm M(\bm M\cdot\bm\Omega) = \\
 &= M_s^2(\bm \Omega -\bm e_z' \Omega_z') \quad.
\end{align}
From the latter equation, one can easily express the angular velocity in terms of magnetization $\bm M$:
\begin{equation}
    \bm \Omega=\frac{\bm M}{M_s^2}\times \dv{\bm M}{t}
+ \bm e_z'\Omega_z' \quad, \label{eq:Omega decomposition}
\end{equation}
which clearly expresses the decomposition of $\bm\Omega$ into components parallel and perpendicular to  $\bm M =M_s\bm e_z'$. Now we assume that the rigid stick is rotationally symmetric along the longitudinal axis $z'$, which means that the inertial tensor $\overleftrightarrow{I}$ is given by:
\begin{equation}
    \overleftrightarrow{I}=\left(\begin{array}{ccc}
       I_\perp'  & 0 & 0 \\
       0  &  I_\perp' & 0 \\
       0  &   0       & I_z'
    \end{array}\right) \quad,
\end{equation}
where the principal moments of inertia $I_\perp',I_z'$ along transverse and longitudinal axes appear. By using the relation between angular momentum $\bm L$ and angular velocity $\bm \Omega$ and remembering the decomposition \eqref{eq:Omega decomposition}, one has:
\begin{equation}
    \bm L= \overleftrightarrow{I}\cdot \bm \Omega= \frac{I_\perp'}{M_s^2}\bm M\times \dv{\bm M}{t}
+ I_z' \bm e_z'\Omega_z' \quad,
\end{equation}
where simple scaling of transverse and longitudinal components occur through the corresponding inertial moments. At this point, following the semiclassical relationship between angular momentum $\bm L$ and magnetic moment $\bm M V$, one can make the identification:
\begin{equation}
    I_z'\Omega_z'=L_z'=-\frac{M_sV}{\gamma_e} \quad,
\end{equation}
which leads to the final expression of the angular momentum:
\begin{equation}
    \bm L = \frac{I_\perp'}{M_s^2}\bm M\times \dv{\bm M}{t}
- \frac{\bm M V}{\gamma_e} \quad.
\end{equation}
Now, considering the dynamics of the rigid body subject to the magnetic torque $\mu_0 \bm M\times \bm H$, one has:
\begin{align}
    \dv{\bm L}{t}& =\bm T = \mu_0V\bm M\times \bm H = \\
    &= \frac{I_\perp'}{M_s^2}\bm M\times \dv[2]{\bm M}{t}
- \frac{V}{\gamma_e}\dv{\bm M}{t} \quad,
\end{align}
which, remembering that $L_z'=-M_z'V/\gamma_e$, leads to the following equation for magnetization dynamics:
\begin{equation}
    \dv{\bm M}{t}=-\gamma\bm M\times \left(\bm H- \frac{I_\perp'}{\mu_0M_s^2 V}\dv[2]{\bm M}{t} \right) \,, \label{eq:torque equation magnetization}
\end{equation}
where $\gamma=\gamma_e \mu_0\approx 2.211\times 10^5\,m A^{-1}s^{-1}$ ($\mu_0$ is the vacuum permeability).
Equation \eqref{eq:torque equation magnetization} clearly relates the occurrence of inertial effects to the presence of a nonzero transverse moment of inertia. In this respect, Gilbert's assumption $I_\perp'=0$ yields the conservative precessional equation; including Gilbert damping then gives the classical LLG equation \eqref{eq:LLG dimensional}. Conversely, a nonzero $I_\perp'$ immediately gives rise to an additional angular momentum relaxation torque, proportional to the second time-derivative of magnetization according to a coefficient that defines the time scale of magnetic inertial effects:
\begin{equation}
    t_\mathrm{in}=\sqrt{\frac{I_\perp'}{\mu_0 M_s^2 V}} \quad,
\end{equation}
which leads to the inertial magnetization dynamics described by Eq.~\eqref{eq:iLLG dimensional} that also includes dissipative effects through the Gilbert damping torque. 

\section{Inertial magnetization dynamics in spatially-inhomogeneous nanomagnets}\label{sec:inertial dynamics inhomogeneous}

\begin{figure}[t]
    \centering
    \includegraphics[width=6cm]{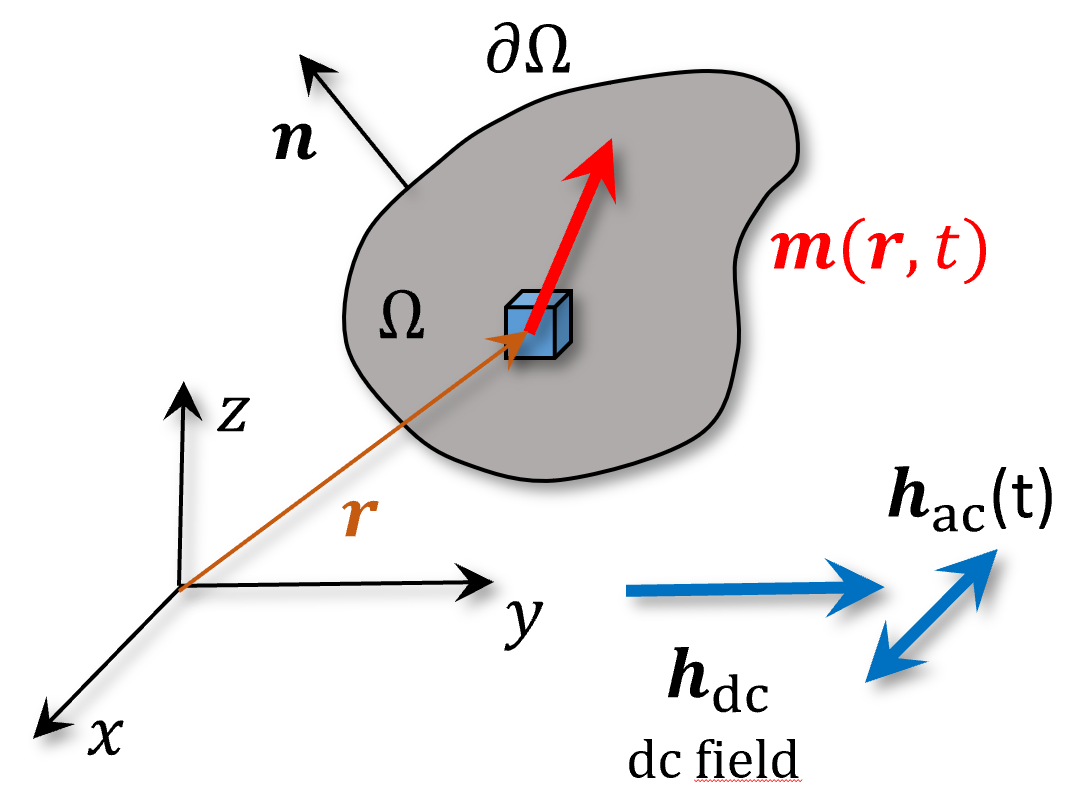}
    \caption{Sketch of confined ferromagnetic particle with spatially-inhomogeneous magnetization occupying region $\Omega$ and subject to external applied fields having constant (dc) and/or time-harmonic (ac) components. In each elementary volume centered at location $\bm r$, $\bm m(\bm r,t)$ denotes the magnetization unit-vector. }
    \label{fig:sketch confined bodies}
\end{figure}

Now we extend the analysis of magnetization dynamics when the macrospin assumption is released, namely the magnetization vector $\bm M(\bm r,t)$ is also a function of the position $\bm r\in \Omega$, with $\Omega$ being the region occupied by the magnetic body, as depicted in Fig.~\ref{fig:sketch confined bodies}. In this respect, $\bm M(\bm r,t) dV$ represents the magnetic moment within the elementary volume $dV$ at location $\bm r$ and time $t$. For the sake of compactness, we introduce dimensionless notation for the relevant quantities, namely magnetization unit-vector $\bm m(\bm r,t)=\bm M/M_s$ and dimensionless effective field $\bm h_\mathrm{eff}(\bm r,t)=\bm H_\mathrm{eff}/M_s$. 

Inertial magnetization dynamics of confined micro- or nanoscale ferromagnets driven by generic external applied fields $\bm h_a$ (also normalized by $M_s$) is driven by the following spatially-inhomogeneous iLLG equation\cite{daquino2023micromagnetic}:
\begin{equation}\label{eq:iLLG equation}
    \pdv{\bm m}{t}=-\bm m\times \left( \bm h_\mathrm{eff}-\alpha \pdv{\bm m}{t}-\xi\pdv[2]{\bm m}{t} \right) \,,
\end{equation}
where time is measured in units of $(\gamma M_s)^{-1}$ (typically of the order of a few picoseconds) and the parameter $\xi=(\gamma M_s t_\mathrm{in})^2$ measures the strength of inertial effects. The effective field $\bm h_\mathrm{eff}$ is computed, according to the micromagnetic theory\cite{brown_micromagnetics_1963,Abert2019}, as the variational derivative of the micromagnetic free energy functional:
\begin{equation} \label{eq:micromagnetic free energy}
   g(\bm m,\bm h_a)\!=\!\frac{1}{V}\!\int_\Omega \!\!\left(\!\frac{\ell_\mathrm{ex}^2}{2}(\nabla \bm m)^2\!+\!f_\mathrm{an}(\bm m)\!-\!\frac{\bm h_m\cdot\bm m}{2}\!-\!\bm h_a\cdot\bm m \right) dV \,,
\end{equation}
which gives
\begin{equation}\label{eq:effective field}
    \bm h_\mathrm{eff}(\bm m,t)=-\fdv{g}{\bm m} = \ell_\mathrm{ex}^2\nabla^2\bm m-\pdv{f_\mathrm{an}}{\bm m}+\bm h_m+\bm h_a\,,
\end{equation}
where $\ell_\mathrm{ex}=\sqrt{2A_\mathrm{ex}/(\mu_0 M_s^2)}$ is the exchange length (typically of the order of several nanometers, with $A_\mathrm{ex}$ being the exchange stiffness constant $\sim 10$ pJ/m), $f_\mathrm{an}(\bm m)$ is the anisotropy energy density, $\bm h_a(\bm r,t)$ is the applied field, and $\bm h_m$ is the magnetostatic (demagnetizing) field that is solution of magnetostatic Maxwell's equations:
\begin{align}
    \nabla&\cdot \bm h_m=-\nabla\cdot \bm m \,\text{ in }\Omega\\
    \nabla&\times\bm h_m=0 \,\text{ in }\Omega \\
    \vu{n}&\cdot[\bm h_m]_{\partial \Omega}=\bm m\cdot \vu{n} \,,
\end{align}
which is readily expressed through the integral operator $\mathcal{N}$ related with the problem Green's function:
\begin{equation}
    \bm h_m(\bm m)=-\mathcal{N}\bm m=\nabla\nabla\cdot\int_\Omega \frac{\bm m(\bm r')}{4\pi\abs{\bm r-\bm r'}}dV \,. \label{eq:magnetostatic field operator}
\end{equation}
When the anisotropy is of uniaxial type, namely $f_\mathrm{an}=\frac{h_\mathrm{an}}{2}[1-(\bm m\cdot \bm e_\mathrm{an})^2]$ (with $h_\mathrm{an}=2K_1/(\mu_0 M_s^2)$ and $\bm e_\mathrm{an}$ being the normalized uniaxial anisotropy field and unit-vector, respectively), the effective field \eqref{eq:effective field} can be conveniently written in operator form as:
\begin{equation}
    \bm h_\mathrm{eff}(\bm m)=-\mathcal{C}\bm m+\bm h_a \,, \label{eq:effective field operator form}
\end{equation}
where $\mathcal{C}=-\ell_\mathrm{ex}^2\nabla^2+\mathcal{N}-h_\mathrm{an}\bm e_\mathrm{an}\otimes\bm e_\mathrm{an}$ is a linear and self-adjoint operator in the appropriate subspace of square-integrable vector fields\cite{brown_micromagnetics_1963}.
The iLLG equation must be complemented with a boundary condition on $\partial \Omega$, which is typically chosen as the natural Neumann one, $\partial\bm m/\partial \bm n=0$ that refers to the absence of surface anisotropy\cite{brown_micromagnetics_1963}.

We observe that, for usual ferromagnetic materials such as cobalt and nickel, both damping $\alpha\sim 10^{-2}$ and intensity of inertia $\xi\sim 10^{-2}$ (when $t_\mathrm{in}\sim$ fractions of ps) are small quantities $\alpha,\xi\ll 1$, meaning that angular momentum relaxation and dissipation act as perturbation of the conservative precessional dynamics.

We also remark that, in the context of inertial dynamics, micromagnetic equilibria are stationary solutions of iLLG Eq.~\eqref{eq:iLLG equation} with zero second-derivative, namely:
\begin{equation}\label{eq:micromagnetic equilibria}
    \pdv{\bm m}{t}=0 \quad,\quad \pdv[2]{\bm m}{t}=0 \quad. 
\end{equation}

\subsection{Inertial Landau-Lifshitz equation}

It is also possible to derive an alternative form for the iLLG Eq.~\eqref{eq:iLLG equation} that resembles the celebrated Landau-Lifshitz (LL) equation augmented with appropriate inertial torque. To this end, we multiply both sides of Eq.~\eqref{eq:iLLG equation} by $\bm m$:
\begin{equation}
\begin{split}
    \bm m \times \pdv{\bm m}{t}&=-\bm m\times (\bm m\times \bm h_\mathrm{eff}) \\ &+ \alpha \left[\bm m\left(\bm m\cdot\pdv{\bm m}{t}\right)-\pdv{\bm m}{t}\left(\bm m\cdot \bm m\right)\right] \\
    &+ \xi \left[\bm m\left(\bm m\cdot\pdv[2]{\bm m}{t}\right)-\pdv[2]{\bm m}{t}\left(\bm m\cdot \bm m\right)\right] = \\
    &=-\bm m\times (\bm m\times \bm h_\mathrm{eff}) \\ & +\alpha \pdv{\bm m}{t} 
    - \xi \left[\bm m\left|\pdv{\bm m}{t}\right|^2+\pdv[2]{\bm m}{t}\right] \,,
    \end{split}
\end{equation}
where we have used the property $|\bm m|^2=1$.
By substituting the right-hand side of the latter equation in the iLLG Eq.~\eqref{eq:iLLG equation}, one has:
\begin{equation}
\begin{split}
      (1+\alpha^2) \pdv{\bm m}{t}&=-\bm m\times \bm h_\mathrm{eff} +\alpha \bm m\times (\bm m\times \bm h_\mathrm{eff})  \\
      &+  \xi\bm m\times \pdv[2]{\bm m}{t}-\alpha\xi\left[\bm m\left|\pdv{\bm m}{t}\right|^2+\pdv[2]{\bm m}{t}\right] \,.
\end{split}
\end{equation}
By neglecting second-order terms $\mathcal{O}(\alpha^2),\mathcal{O}(\alpha\xi)$, one ends up with the following inertial Landau-Lifshitz (iLL) equation:
\begin{equation}
    \pdv{\bm m}{t}=-\bm m\times \left(\bm h_\mathrm{eff}-\alpha \bm m\times \bm h_\mathrm{eff} + \xi \pdv[2]{\bm m}{t}\right) \,. \label{eq:iLL equation}
\end{equation}
The iLL Eq.~\eqref{eq:iLL equation} is equivalent to the iLLG Eq.~\eqref{eq:iLLG equation} for small damping and small intensity of inertial effects in analogy with what occurs for the classical Landau-Lifshitz and Landau-Lifshitz-Gilbert equations.

\subsection{Dynamical system form and qualitative properties of iLLG dynamics}

\begin{figure}[t]
    \centering
    \includegraphics[width=3cm]{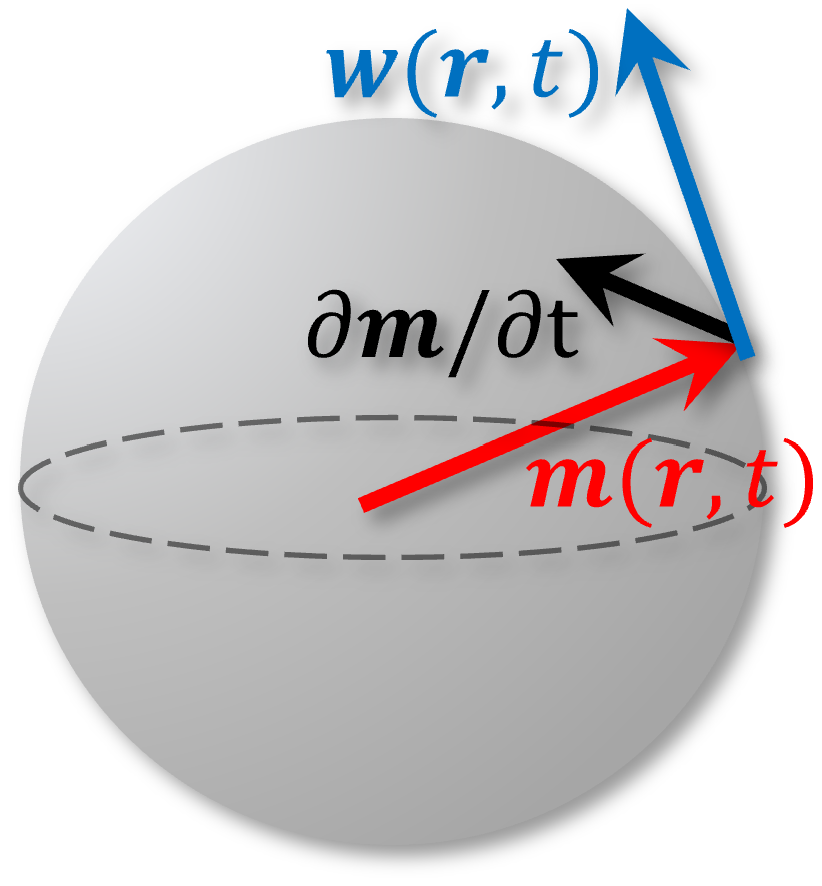}
    \caption{Formal mechanical analogy for the introduction of variable $\bm w=\bm m\times \pdv{\bm m}{t}$ as the 'angular momentum' of a point-particle of unitary mass, position vector $\bm m$ and velocity $\pdv{\bm m}{t}$. }
    \label{fig:formal mechanical sketch}
\end{figure}

The iLLG Eq.~\eqref{eq:iLLG equation} has implicit form since the unknown vector field $\bm m$ appears both in the left and right sides. An  alternative explicit form can be derived\cite{daquino2024midpoint} by using a two-vector-fields formulation. To this end, besides magnetization $\bm m$, we introduce the vector field $\bm w(\bm r,t)$
\begin{equation}\label{eq:w definition}
    \bm w=\bm m\times \pdv{\bm m}{t} \quad\Rightarrow\quad \pdv{\bm w}{t}=\bm m\times\pdv[2]{\bm m}{t} \,.
\end{equation}
Within a purely formal mechanical analogy, the newly defined variable $\bm w(\bm r,t)$  resembles the 'angular momentum' of a point-particle of unitary mass, position vector $\bm m$ and velocity $\pdv{\bm m}{t}$, as it is sketched in Fig.~\ref{fig:formal mechanical sketch}.

By taking the cross product with $\bm m$ on both sides
of the first of Eqs.~\eqref{eq:w definition}, and then
using Eqs.~\eqref{eq:w definition} to rewrite the inertial
terms in the iLLG Eq.~\eqref{eq:iLLG equation}, one obtains
the coupled evolution equations\cite{daquino2024midpoint}:
\begin{align}
    \pdv{\bm m}{t}&=\bm w\times\bm m \,, \label{eq:dm_dt} \\ 
    \xi\pdv{\bm w}{t}&=-\bm m\times \bm w-\alpha\bm w+\bm m\times \bm h_\mathrm{eff} \,. \label{eq:dw_dt}
\end{align}
Equations \eqref{eq:dm_dt}-\eqref{eq:dw_dt}, complemented with natural boundary conditions on $\partial \Omega$, define a spatially-distributed dynamical system  with the two independent state-variables $\bm m$ and $\bm w$, which is equivalent to the iLLG Eq.~\eqref{eq:iLLG equation} when Eq.~\eqref{eq:w definition} holds at the initial time $t_0$. The micromagnetic equilibria are then given by the stationary solutions of Eqs.~\eqref{eq:dm_dt}-\eqref{eq:dw_dt}. It is apparent that the iLLG dynamics requires twice as many degrees of freedom compared to the classical LLG for its description.

Important qualitative properties of the iLLG dynamics can be straightforwardly derived\cite{daquino2024midpoint} using the above formulation: 
\begin{itemize}
\item \textbf{Property \#1:} magnetization $\bm m$ evolves on the unit-sphere, that is $\forall \bm r\in \Omega\,,\,t\geq t_0$
\begin{equation}\label{eq:constant amplitude}
     \abs{\bm m(\bm r,t)}=1 \,.
\end{equation}
It follows immediately from Eq.~\eqref{eq:dm_dt} dot-multiplied by $\bm m$.

\item \textbf{Property \#2:} the 'angular momentum' variable $\bm w$ relaxes towards the plane tangent to the unit-sphere with decay constant $\xi/\alpha$, that is $\forall\bm r\in\Omega, t\geq t_0 $:
\begin{equation}\label{eq: w relaxation}
    \bm w(\bm r,t)\cdot\bm m(\bm r,t)= \bm w(\bm r,t_0)\cdot\bm m(\bm r,t_0)e^{-\frac{\alpha}{\xi}(t-t_0)} \,.
\end{equation}
This result follows by dot-multiplying Eq.~\eqref{eq:dm_dt} by
$\bm w$, dot-multiplying Eq.~\eqref{eq:dw_dt}, divided by $\xi$,
by $\bm m$, and summing the resulting equations.

\item \textbf{Property \#3:} The iLLG dynamics fulfills a generalized free energy balance, that is $\forall t\geq t_0$:
\begin{equation}\label{eq:energy balance}
    \dv{\tilde{g}}{t}=-\frac{1}{V}\int_\Omega \left(\bm m\cdot\pdv{\bm h_a}{t}+\alpha\left|\pdv{\bm m}{t}\right|^2\right) dV \,,
\end{equation}
where the functional $\tilde{g}$ is given by the sum of the micromagnetic free energy \eqref{eq:micromagnetic free energy} with an additional term pertaining to inertial effects:
\begin{align}
    \tilde{g}(\bm m,\bm w,\bm h_a)&=g(\bm m,\bm h_a) + \frac{1}{V}\int_\Omega \frac{\xi}{2}\left|\bm w\right|^2 dV \,, \label{eq:generalized free energy}
\end{align}
where $|\bm w|$ can  be replaced with $|\partial\bm m/\partial t|$ when Eq.~\eqref{eq:w definition} holds. In Eq.~\eqref{eq:generalized free energy}, the micromagnetic free energy $g$ plays the role of a 'potential' energy while the additional term in Eq.~\eqref{eq:generalized free energy} resembles a 'kinetic' energy, since it is a function of the square of 'velocity'. 
In particular, Property \#3 tells that, for constant-in-time applied field, the iLLG dynamics admits a Lyapunov structure\cite{daquino2024midpoint}, namely the generalized free energy $\tilde{g}$ is a decreasing function of time, implying that the dynamics evolves toward stable equilibrium configurations.
Furthermore, when the dissipation is absent (i.e. $\alpha=0$), the generalized free energy is conserved.
\end{itemize}

The above three qualitative properties express general constraints for the inertial magnetization dynamics. Specifically, Properties \#1 and \#2 establish unit-sphere constraint for the magnetization $\bm m$ and relaxation for the 'angular momentum' variable $\bm w$, respectively. In addition, Property \#3 establishes the way energy can be pumped into the magnetic system through time-varying external fields and the way it is dissipated owing to the intrinsic damping mechanism. 

Besides their physical significance, these qualitative
properties provide fundamental criteria for assessing
the reliability of numerical schemes for inertial
magnetization dynamics. In particular, their preservation
at the discrete level will play an important role in the
geometric integration methods discussed in Sec.~\ref{sec:numerical}.

It should also be emphasized that the iLLG equation
considered here describes transverse magnetization
dynamics under the assumption of constant magnetization
amplitude. Consequently, it is not intended to describe
ultrafast demagnetization processes involving substantial
longitudinal relaxation of magnetization, nor does it
explicitly account for nonequilibrium electronic and
lattice degrees of freedom. The inertial term provides
a phenomenological description of angular-momentum
relaxation within this framework, whose range of applicability ultimately depends on the
material and on the characteristic temporal and spatial
scales of the investigated process.

\section{Modelling inertial spin-wave dynamics}\label{sec:modelling inertial SW dynamics}



\subsection{Overview of literature}


The extension of conventional spin-wave theory to the
inertial regime has stimulated considerable research
on the spectral and propagation properties of nutational
excitations. In particular, the inclusion of magnetic
inertia modifies the ordinary precessional branches
and introduces additional high-frequency collective
modes, commonly referred to as nutational spin waves
or inertial magnons.

One of the first theoretical investigations of the
consequences of magnetic inertia for spatially
inhomogeneous magnetization dynamics was carried out
by Kikuchi and Tatara~\cite{kikuchi2015spin}, who
derived an effective spin inertia in metallic
ferromagnets and analyzed its influence on spin-wave
and domain-wall dynamics.
Subsequently, Makhfudz et
al.~\cite{makhfudz2020nutation} investigated the
emergence of collective nutation waves in
exchange-coupled ferromagnetic spin systems,
discussing their characteristic high frequencies
and propagation properties.

The frequency-domain response of inertial
ferromagnets was investigated by Cherkasskii et
al.~\cite{cherkasskii2020nutation}, who derived
analytical expressions for the resonance frequencies
and linewidths of precessional and nutational
modes from the magnetic susceptibility tensor.
The extension to spatially inhomogeneous dynamics
was addressed by the same authors in
Ref.~\cite{cherkasskii2021dispersion}, where
the dispersion relations of nutational surface
spin waves in ferromagnetic thin films were
investigated within the magnetostatic approximation.
In particular, the occurrence of backward
nutational spin waves and the inertial redshift
of conventional magnetostatic spin-wave branches
were demonstrated.

A complementary analysis was developed by
Lomonosov et al.~\cite{lomonosov2021anatomy},
who investigated the dispersion relations,
eigenfrequencies and quality factors of inertial
magnons in ferromagnetic nanostructures with
different geometries and magnetic anisotropies.
Their results also highlighted the possibility
of exciting nutational magnons through
magnetoelastic coupling.
The influence of magnetic anisotropy on
precessional and nutational resonances was
further analyzed by Cherkasskii et
al.~\cite{cherkasskii2022theory}, who developed
a generalized resonance theory for anisotropic
ferromagnets, providing analytical expressions
for the inertial frequency shifts.

The general theory of inertial spin waves
was subsequently extended by Mondal and
R\'ozsa~\cite{mondal2022inertial} to
ferromagnets and two-sublattice antiferromagnets.
Their analysis showed how the coupling between
precessional and nutational excitations modifies
the spin-wave frequencies, group velocities,
effective gyromagnetic ratios and damping
parameters.
In parallel, Titov et al.~\cite{titov2022nutation}
investigated nutational spin-wave propagation
by combining the inertial magnetization dynamics
with Maxwell's equations, deriving dispersion
relations for different propagation directions
and analyzing the hybridization between
electromagnetic and nutational excitations.

More recently, Ghosh et al.~\cite{ghosh2026engineering}
extended inertial spin-wave theory to the case of a general
magnetic inertia tensor, investigating the influence of
isotropic, anisotropic and chiral inertial contributions
on the magnon spectrum of two-sublattice ferromagnets.
Their analysis revealed additional possibilities for
controlling the spin-wave band structure and inducing
nonreciprocal propagation through tensorial magnetic
inertia, including chiral inertial effects.

The possibility of exciting ultrashort inertial
spin waves in realistic ferromagnetic
nanostructures was investigated by
d'Aquino et al.~\cite{daquino2023micromagnetic},
who developed an analytical description of
exchange-dominated inertial spin waves and
performed spatially resolved micromagnetic
simulations of confined ferromagnetic nanodots.
Their analysis predicted a finite limiting
spin-wave propagation velocity and demonstrated
the excitation of nutational spin waves with
wavelengths of the order of tens of nanometers
by THz electromagnetic fields.
The influence of spin-transfer torques on the
propagation and dissipation of inertial spin
waves was subsequently investigated by
He and Cherkasskii~\cite{he2024temporal},
who analyzed the dependence of temporal
attenuation and spatial decay lengths on
magnetic inertia and electric current.

More recently, the theory of inertial spin
waves has been extended to magnetic systems
with noncollinear equilibrium configurations.
In particular, Cherkasskii, Mondal and
R\'ozsa~\cite{cherkasskii2024inertial}
developed a general linear spin-wave framework
including magnetic inertia and applied it
to planar and conical spin spirals, revealing
important modifications of the nutational
dispersion branches associated with the
underlying magnetic structure.
The high-frequency magnetic susceptibility
in the inertial regime was further investigated
by Wiemeler et al.~\cite{wiemeler2025high},
providing additional theoretical insight
into the spectral response of ferromagnets
at sub-THz frequencies.

An interesting recent development concerns
the interplay between magnetic inertia
and the topology of collective spin
excitations.
Ghosh et al.~\cite{ghosh2026spin}
showed theoretically that the coupling
between precessional and nutational magnon
branches can produce topologically
nontrivial band gaps and chiral edge
states in suitable magnetic lattices.
These results suggest additional possibilities
for controlling high-frequency spin-wave
propagation through the combined effects
of magnetic inertia and magnetic interactions.

Another important development concerns the
influence of geometrical curvature and torsion
on inertial spin-wave dynamics.
In particular, d'Aquino and
Hertel~\cite{dAquino2025}
demonstrated theoretically that twisted
magnetic nanostrips exhibit nonreciprocal
precessional and nutational spin-wave
propagation, with opposite asymmetries
in the corresponding dispersion branches, and highlighted the interplay between chirality and confinement in nanostrips with non-trivial topology.

In the following sections, we focus on the
analytical modelling of inertial spin waves
in ferromagnetic systems with uniform
equilibrium magnetization, considering
macrospin particles and extended magnetic
thin films in the classical backward-volume,
Damon--Eschbach and forward-volume
configurations.
We then discuss the extension to curved
magnetic nanostrips, where geometrical
curvature and torsion introduce additional
effects on the dispersion relations,
including nonreciprocal propagation.
The subsequent section will address
spatially resolved micromagnetic simulation
methods and representative applications
to the excitation and propagation of
inertial spin waves in confined
ferromagnetic nanostructures.

\subsection{General problem statement}

Here we consider the small-amplitude magnetization dynamics that gives rise to the emergence of inertial spin waves oscillating in the ferromagnet. We assume that the ferromagnetic body is uniformly-magnetized at equilibrium along a given direction $\bm m_0$, subject to a time-varying field $\bm h_a(t)=\bm h_\mathrm{dc}+\delta\bm h_\mathrm{ac}(t)$ such that $|\delta\bm h_\mathrm{ac}(t)|\ll|\bm h_\mathrm{dc}|$, and study the dynamics of small deviations $\delta\bm m(\bm r,t)$ from the equilibrium magnetization:
\begin{equation}\label{eq:magnetization deviation}
    \bm m(\bm r,t)=\bm m_0+\delta\bm m(\bm r,t) \,.
\end{equation}
To this end, the iLLG Eq.~\eqref{eq:iLLG equation} is linearized around $\bm m_0$, giving rise to the following equation:
\begin{equation}
    \pdv{\delta\bm m}{t}=\bm m_0 \times \left(-\delta\bm h_\mathrm{eff}+h_0\delta\bm m+\alpha\pdv{\delta\bm m}{t} +\xi\pdv[2]{\delta\bm m}{t} -\delta \bm h_\mathrm{ac}\right) \,, \label{eq:linearized iLLG}
\end{equation}
from which it is apparent that the dynamics of deviations $\delta \bm m(\bm r,t)$ will occur on the plane transverse to the equilibrium $\bm m_0$ owing to the fact that, to first order in $\delta\bm m$, one has:
\begin{equation}
    \delta \bm m\cdot\bm m_0 =0 \quad, \label{eq:transverse deviation}
\end{equation}
which is in turn a consequence of Property \#1 (see Eq.~\eqref{eq:constant amplitude}) upon substitution of the decomposition \eqref{eq:magnetization deviation}.
In Eq.~\eqref{eq:linearized iLLG}, $\delta\bm h_\mathrm{eff}$ represents the effective field produced by the linear terms in the effective field \eqref{eq:effective field} acting on magnetization perturbation $\delta\bm m$ and $h_0$ is the projection of the effective field produced by the equilibrium magnetization on itself, that is $h_0=\bm h_\mathrm{eff}(\bm m_0)\cdot \bm m_0$, which includes the contribution from dc applied field and anisotropy when present.

In the following, we will briefly summarize the analytical results holding for inertial spin waves in relevant situations such as macrospin particles, flat (infinite) thin films and curved nanostrips, which can be instrumental to design and/or interpret possible experiments.

\subsection{Frequency response of macrospin particles}

We analyze the frequency response of a spatially-homogeneous thin film magnetized along the equilibrium orientation $\bm m_0$ and subject to an ac field with small amplitude. To this end, let us assume that the applied field is decomposed into a nonzero constant bias field and a time-harmonic component $\bm h_a(t)=\bm h_\mathrm{dc}+\delta \bm h_\mathrm{ac}(t)\,\,, \quad |\delta\bm h_\mathrm{ac}|\ll|\bm h_\mathrm{dc}|$. We also assume that the anisotropy is of uniaxial type, that is $f_\mathrm{an}(\bm m)=\frac{K_1}{\mu_0 M_s^2}[1-(\bm m\cdot \bm e_\mathrm{an})^2]$ with $e_\mathrm{an}$ being the easy (resp. hard) direction if the anisotropy constant $K_1$ is positive (resp. negative). Consequently, under the macrospin approximation, the free energy \eqref{eq:micromagnetic free energy} has the simple form :
\begin{equation}\label{eq:free energy macrospin}
    g(\bm m, \bm h_a)=\frac{1}{2} D_x m_x^2 + \frac{1}{2} D_y m_y^2 +\frac{1}{2} D_z m_z^2-\bm m\cdot \bm h_a \quad,
\end{equation}
where $D_x,D_y,D_z$ are effective demagnetizing factors taking into account shape and crystalline anisotropy. In the specific case considered, we have:
\begin{equation}
    D_x=N_x-h_\mathrm{an}\,,\,D_y=N_y \,, D_z=N_z \,,
\end{equation}
where $h_\mathrm{an}=2K_1/(\mu_0 M_s^2)$, $N_x,N_y,N_z$ are the principal demagnetizing factors of the demagnetizing tensor arising from Eq.~\eqref{eq:magnetostatic field operator} under the macrospin assumption, such that $N_x+N_y+N_z=1$ owing to the unit-trace property of demagnetizing operator\cite{Moskowitz1966}.

Then, the linearized iLLG Eq.~\eqref{eq:linearized iLLG} reads as:
\begin{equation}\label{eq:linear iLLG}
    \frac{d \delta \bm m}{d t}=\bm m_0 \times \left[(\overleftrightarrow{D}+h_0\overleftrightarrow{I}) \cdot\delta \bm m + \alpha\frac{d \delta \bm m}{d t} + \xi \frac{d^2 \delta \bm m}{d t^2} - \delta\bm h_\mathrm{ac}\right] ,
\end{equation}
where $\overleftrightarrow{D}$ denotes the diagonal tensor $\overleftrightarrow{D}=\text{diag}[D_x,D_y,D_z]$, $h_0=\bm h_\mathrm{eff}(\bm m_0)\cdot \bm m_0 =(-\overleftrightarrow{D}\cdot \bm m_0 +\bm h_\mathrm{dc}) \cdot \bm m_0$ is the projection of the equilibrium effective field on equilibrium and  $\overleftrightarrow{I}$ is the identity tensor.

According to Eq.~\eqref{eq:transverse deviation},  we can consider only the dynamics of the component $\delta \bm m_\perp$ of $\delta \bm m$ living in the plane perpendicular to the equilibrium $\bm m_0$. 
As a consequence, we will deal with vectors having only two components associated with axes transverse to the equilibrium $\bm m_0$. 
We express the cross product in operator form $\bm v\times \bm w=\Lambda(\bm v)\cdot\bm w$ using the skew-symmetric operator $\Lambda$ that is invertible  (such that $\Lambda\cdot\Lambda=-\overleftrightarrow{I}$) when restricted to the plane orthogonal to $\bm m_0$ and write both $\delta \bm m_\perp,\delta\bm h_\mathrm{ac}$ using complex (phasor) domain as $\delta \bm m_\perp(t)=\delta  \tilde{\bm m} e^{i\omega t} \,\,,\,\, \delta\bm h_\mathrm{ac}=\tilde{\delta\bm h}_\mathrm{ac} e^{i\omega t}$.
By using these formulas in Eq.~\eqref{eq:linear iLLG}, one ends up with:
\begin{equation}\label{eq:iLLG an freq resp}
    \delta \tilde{\bm m}= \left[\mathcal{P}_\perp [i\omega\Lambda(\bm m_0)]+\overleftrightarrow{D}+(h_0-\xi\omega^2+i\omega\alpha)\overleftrightarrow{I}\right]^{-1} \!\!\tilde{\delta \bm h}_\mathrm{ac} \,\,,
\end{equation}
where $\mathcal{P}_\perp$ denotes the projection operator on the plane transverse to $\bm m_0$ and the inverse of the matrix in square brackets defines the magnetic susceptibility tensor $\overleftrightarrow{\chi}(\omega)$.
When referred to the axes transverse to $\bm m_0$, $\overleftrightarrow{\chi}(\omega)$ becomes a $2\times 2$ matrix which can be easily computed.

The resonance frequencies associated with the above linear dynamical system are associated with nontrivial solutions of the unforced dynamics  (i.e. Eq.~\eqref{eq:iLLG an freq resp} under zero ac field) such that 
\begin{equation}\label{eq:det inv_Chi_omega}
\det[\overleftrightarrow{\chi}(\omega)^{-1}]=0 \,.     
\end{equation}

For a specific choice of $\bm m_0$ and $\bm h_\mathrm{dc}$ aligned, for instance, with the $x$ axis, Eq.~\eqref{eq:det inv_Chi_omega} can be put in a convenient form (see Eq.~\eqref{eq:gen characteristic poly sym} in Appendix) by 
defining $\omega_{11}=\omega_{0y}=D_y-D_x+h_\mathrm{dc}$, $\omega_{22}=\omega_{0z}=D_z-D_x+h_\mathrm{dc}$ and using appropriate perturbation theory leading to the following resonance frequencies (computed in the conservative case when $\alpha=0$):
\begin{align}
    \omega_N &\approx \pm \sqrt{ \frac{\sqrt{2\xi(\omega_{0y}+\omega_{0z})+1} + \xi(\omega_{0y}+\omega_{0z})+1}{2\xi^2} } \,\,,\label{eq:resonance frequencies nut} \\
    \omega_P&\approx \pm\frac{\omega_K}{\sqrt{1+\xi(\omega_{0y}+\omega_{0z}})} \,\,,\,\,\omega_K=\sqrt{\omega_{0y}\omega_{0z}} \label{eq:resonance frequencies prec}\,\,,
\end{align}
where $\omega_K=\sqrt{(D_y-D_x+h_\mathrm{dc})(D_z-D_x+h_\mathrm{dc})}$ is the classical Kittel ferromagnetic resonance (FMR) frequency. 
Equation \eqref{eq:resonance frequencies prec} describes the influence of inertial effects on the FMR frequency, which results in a redshift\cite{Ciornei2011magnetization} compared to the non-inertial situation $\xi=0$, while the formula \eqref{eq:resonance frequencies nut} gives the nutation resonance frequency. The former is typically in the GHz range, being $\omega_K\sim\mathcal{O}(1)$, while the latter falls within the THz range, as it can be inferred performing the following expansion for $\xi\ll1$:
\begin{equation}
    \omega_N \overset{\xi\ll 1}{\approx} \frac{1}{\xi}+\frac{\omega_{0y}+\omega_{0z}}{2} \quad,
\end{equation}
and observing that, since $\xi\sim 10^{-2}$, $\omega_N$ is at least two orders of magnitude larger than $\omega_K$ (conversion to Hz is obtained by multiplying
$\omega_K$ and $\omega_N$ by the dimensional factor
$\gamma M_s/(2\pi)$).

We observe that the above formulas take into account the dependence on the external bias field through the parameters $\omega_{0y},\omega_{0z}$.
Using the approach reported in appendix \ref{sec:appendix}, it is also possible to determine closed-form expressions for the full width at half maximum (FWHM) linewidths:
\begin{align}
2\Delta\omega_P
&\approx
\alpha(\omega_{0y}+\omega_{0z})
\left[
1-\frac{3}{2}\xi(\omega_{0y}+\omega_{0z})
\right], \label{eq:resonance linewidths prec} \\
2\Delta \omega_N &\approx \frac{\alpha}{\xi}\left[1+\frac{1}{\sqrt{2\xi(\omega_{0y}+\omega_{0z})+1}} \right] \,\,. \label{eq:resonance linewidths nut}
\end{align}
Here and in the following, $\Delta\omega$ denotes the
half-width at half maximum (HWHM), so that the corresponding
full-width at half maximum is $2\Delta\omega$.

For illustrative purposes, we consider an infinite permalloy thin film ($D_x=0, D_y=0, D_z=1)$ with material parameters\cite{Neeraj2021inertial}: damping $\alpha=0.023$, saturation magnetization such that $\mu_0 M_s=0.93~$T, inertial time scale $t_\mathrm{in}=1.26$ ps and applied field along $x$ equal to 100~mT. With these values, $\omega_K\approx 0.3451 \rightarrow f_K=8.9871$ GHz and Eq.~\eqref{eq:resonance frequencies nut} gives $\omega_N\approx 24.1328 \rightarrow f_N=628.4782$ GHz. We notice that, as implied by formula \eqref{eq:resonance frequencies prec}, the FMR frequency $\omega_P$ is always red-shifted\cite{Ciornei2011magnetization} by the magnetic inertia compared to $\omega_K$ (in fact, $\omega_P=\omega_K$ when $\xi=0$) and for the present case gives 8.7636 GHz corresponding to a $2.5\%$ redshift relative to the
non-inertial case.

Concerning the precessional FWHM linewidth, Eq.~\eqref{eq:resonance linewidths prec} shows that the presence of inertia $\xi\neq 0$ produces a reduction compared to the non-inertial case.
For the chosen parameters, Eq.~\eqref{eq:resonance linewidths prec} gives
$2\Delta\omega_P=0.0258$, corresponding to
$2\Delta f_P=671.4$ MHz, compared with
$2\Delta\omega_K=0.0279$ and
$2\Delta f_K=727.8$ MHz in the non-inertial case,
namely a reduction of approximately $8\%$.
These results summarize the main signatures of magnetic
inertia in the spatially uniform response: the emergence
of a high-frequency nutational resonance and the redshift
and linewidth reduction of the conventional precessional
resonance. In the following, we extend this picture to
spatially inhomogeneous excitations and inertial spin-wave
propagation in magnetic thin films.

\subsection{Spin waves in flat extended thin films}

\begin{figure*}[t]
    \centering
    \includegraphics[width=17cm]{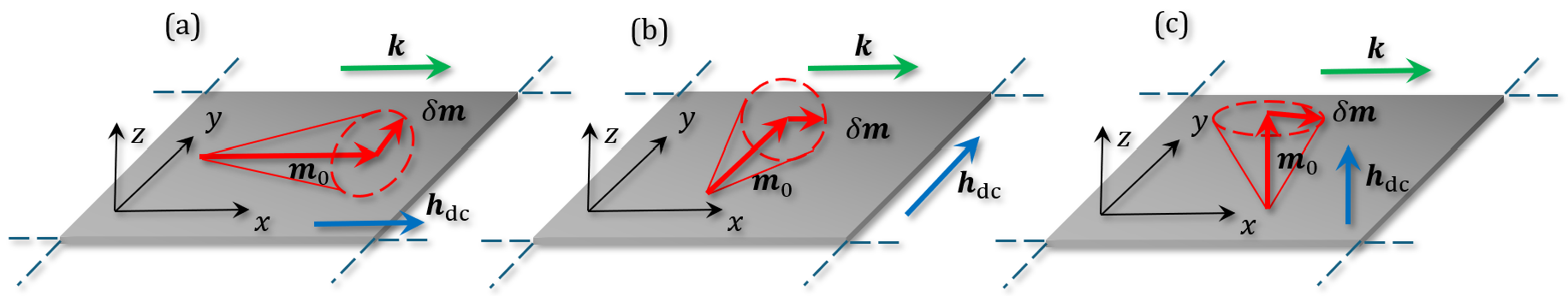}
    \caption{Sketch of different geometries for inertial spin-wave excitation in flat extended magnetic thin films: (a) backward volume; (b) Damon-Eschbach; (c) forward volume. In all panels, $\bm m_0$ denotes the equilibrium (static) magnetization, $\delta\bm m$ the spin-wave field representing the dynamics deviation from equilibrium oscillating in the transverse plane, $\bm h_\mathrm{dc}$ is the static (dc) applied field, $\bm k$ is the spin-wave wavevector.}
    \label{fig:SW geometries flat film}
\end{figure*}

In this section, we briefly summarize the main results concerning inertial spin waves dynamics occurring in extended magnetic thin films. To this end, we assume translational invariance along the in-plane directions $x,y$ and thickness $d$ along the out-of-plane $z$ direction. 
We will consider the classical geometries for spin-wave
propagation in flat magnetic thin films, extensively
investigated both theoretically and experimentally
\cite{Carlotti2002,Kalinikos1986}, namely backward volume
(BV), Damon--Eshbach (DE) and forward volume (FV), as
illustrated in Fig.~\ref{fig:SW geometries flat film}. In all cases,
spin-wave propagation is assumed to occur along the
direction of the wavevector $\bm k$.

The starting point is the linearized iLLG equation \eqref{eq:linearized iLLG} rewritten in the Fourier domain:
\begin{equation}\label{eq:linearized iLLG fourier}
    i\omega \delta\hat{\bm m} = \bm m_0\times \left(-\delta\hat{\bm h}_\mathrm{eff}+h_0 \delta\hat{\bm m} + i\omega\alpha\delta\hat{\bm m}-\xi\omega^2\delta\hat{\bm m}-\delta\hat{\bm h}_\mathrm{ac}\right) \,,
\end{equation}
where $\delta\hat{\bm m}(\bm k,\omega), \delta\hat{\bm h}_\mathrm{ac}(\bm k,\omega)$ are the Fourier Transforms of $\delta\bm m(\bm r,t), \delta\bm h_\mathrm{ac}(\bm r,t)$, respectively computed using the following definition:
\begin{equation}\label{eq:fourier transform}
    \hat{\bm v}(\bm k,\omega)=\int \bm v(\bm r,t)e^{i(\bm k\cdot\bm r-\omega t)} \,d\bm r\, dt \,.
\end{equation}
The effective field $\delta \bm h_\mathrm{eff}[\delta\bm m]$ defines a linear integral-differential operator acting on the magnetization deviation $\delta\bm m(\bm r,t)$ according to Eqs.~\eqref{eq:effective field}-\eqref{eq:magnetostatic field operator}. In this respect, assuming uniaxial anisotropy, that is $f_\mathrm{an}(\bm m)=\frac{h_\mathrm{an}}{2}[1-(\bm m\cdot \bm e_\mathrm{an})^2]$ with $e_\mathrm{an}$ being the easy (resp. hard) direction if the anisotropy field $h_\mathrm{an}$ is positive (resp. negative), the linear effective field $\delta\bm h_\mathrm{eff}$ becomes:
\begin{equation}
    \delta\bm h_\mathrm{eff}=[\ell_\mathrm{ex}^2\nabla^2-\mathcal{N}+h_\mathrm{an}\bm e_\mathrm{an}\otimes \bm e_\mathrm{an}]\delta \bm m =-\mathcal{C}\delta\bm m\,.\label{eq:effective field operator}
\end{equation}
The above operator $\mathcal{C}$ can be conveniently written in the Fourier domain as a second-order tensor:
\begin{equation}
    -\overleftrightarrow{C}(\bm k)=-\ell_\mathrm{ex}^2k^2 \overleftrightarrow{I}-\overleftrightarrow{N}(\bm k)+h_\mathrm{an}\bm e_\mathrm{an}\otimes \bm e_\mathrm{an} \,,\label{eq:effective field operator fourier}
\end{equation}
where $\overleftrightarrow{I}$ is the identity tensor, $k=\abs{\bm k}$ and the demagnetizing operator $\mathcal{N}$, averaged along the film thickness, has become a local demagnetizing tensor $\overleftrightarrow{N}(\bm k)$ which is a function of the wavevector. For flat thin films, the demagnetizing tensor can be expressed in the compact form\cite{Kalinikos1986,BMS2009}:
\begin{equation}\label{eq:demagnetizing tensor fourier}
    \overleftrightarrow{N}(\bm k)=\frac{\bm k\otimes\bm k}{k^2}(1-s_k)+\bm e_z\otimes\bm e_z s_k \,,
\end{equation}
where $\bm k=k_x\bm e_x + k_y\bm e_y$ is in-plane and the effective wavenumber-dependent demagnetizing factor is: 
\begin{equation}
s_k=\frac{1-e^{-kd}}{kd} \,. \label{eq:demag fact Fourier}    
\end{equation}

By using Eq.~\eqref{eq:effective field operator fourier} in the linearized iLLG Eq.~\eqref{eq:linearized iLLG fourier} and multiplying both sides times $-\Lambda(\bm m_0)$, one ends up with the following equation:
\begin{align}
    \delta&\hat{\bm m}=\overleftrightarrow{\chi}(\bm k,\omega)\cdot\delta\hat{\bm h}_\mathrm{ac}=\Big\{\mathcal{P}_\perp[i\omega\Lambda(\bm m_0)+\ell_\mathrm{ex}^2k^2\overleftrightarrow{I}+\overleftrightarrow{N}(\bm k) \\
    &-h_\mathrm{an} \bm e_\mathrm{an}\otimes \bm e_\mathrm{an} +
    (h_0-\xi\omega^2+i\omega\alpha)\overleftrightarrow{I}]  \Big \}^{-1}\!\!\!\! \cdot \delta\hat{\bm h}_\mathrm{ac} \,, \label{eq:iLLG SW freq resp}
\end{align}
where, once again, the inverse of the quantity within the brackets defines the susceptibility tensor $\overleftrightarrow{\chi}(\bm k,\omega)$ that exhibits dependence on the wavenumber $k$ besides that on the frequency $\omega$. When referred to the axes transverse to $\bm m_0$, the tensor $\overleftrightarrow{\chi}(\bm k,\omega)$ becomes a $2\times 2$ matrix which can be easily computed.

The resonance frequencies associated with linear dynamical system \eqref{eq:linearized iLLG fourier} are associated with nontrivial solutions of the unforced dynamics  (i.e. Eq.~\eqref{eq:iLLG SW freq resp} under zero ac field) such that 
\begin{equation}
\det[\overleftrightarrow{\chi}(\bm k,\omega)^{-1}]=0 \,, \label{eq:det inv_Chi zero SW} 
\end{equation}
which defines the spin-wave dispersion relation $\omega(\bm k)$ for zero damping ($\alpha=0$). The presence of dissipation implies the occurrence of nonzero spectral (full-width at half-maximum (FWHM)) linewidths that are also important for the experiments. A convenient approach based on perturbation theory is developed to obtain simple closed-form expressions for roots $\pm\omega_N+i\Delta\omega_N,\pm\omega_P+i\Delta\omega_P$ of the fourth-degree polynomial \eqref{eq:det inv_Chi zero}, where the real parts correspond to the resonance frequencies and the imaginary parts yield the FWHM linewidths  $2\Delta\omega_N,2\Delta\omega_P$, both being functions of the wavenumber $\bm k$ and of geometrical and material parameters. 

It is worth remarking that, unlike classical LLG
dynamics, in the inertial case the dispersion relation has multiple branches, one pair associated with nutational spin waves at frequency $\pm\omega_N$ that are an intrinsic consequence of magnetic inertia and another pair associated with precessional spin waves at frequency $\pm\omega_P$, also affected by the presence of inertia that slightly modifies the classical dispersion relation.
The three different excitation geometries for inertial spin waves are specifically addressed in the following paragraphs. 

\subsubsection{Backward volume (BV) inertial spin waves}

\begin{figure}
    \centering
    \includegraphics[width=7.5cm]{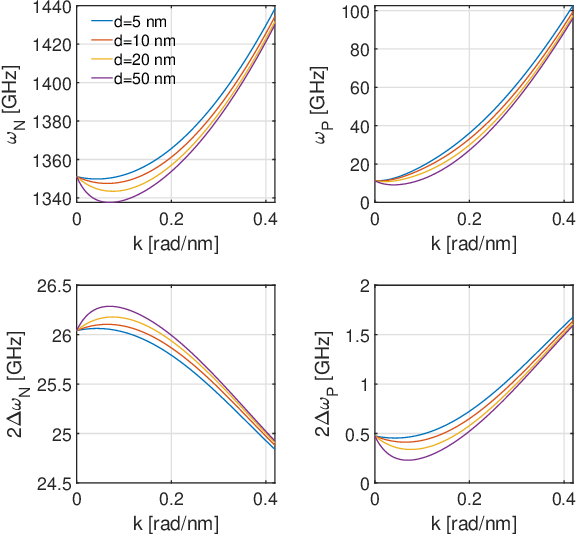}
    \caption{Dispersion relations and FWHM linewidths for nutational and precessional BV inertial spin waves in cobalt thin films as a function of film thickness $d$. Values of parameters: $\gamma=2.211\times 10^5$ m$\,$A$^{-1}\,$s$^{-1}$, $\mu_0 M_s=1.6$ T, $A=13$ pJ/m ($l_\mathrm{ex}=3.57$ nm), $t_\mathrm{in}=0.653$ ps $(\xi=0.0338)$ and $\alpha=0.01$. Upper panels show nutational and precessional dispersion relations; lower panels show the corresponding FWHM linewidths.}
    \label{fig:BV_iSW_dispersion}
\end{figure}

In this BV configuration, the wavevector $\bm k$, the equilibrium $\bm m_0$ and the static applied field $\bm h_\mathrm{dc}$ are parallel (see Fig.~\ref{fig:SW geometries flat film}(a)). We assume, for the sake of simplicity, that $\bm k=k\bm e_x, \bm m_0=\bm e_\mathrm{an}=\bm e_x, \bm h_\mathrm{dc}=h_{ax}\bm e_x$, which imply $h_0=h_{ax}+h_\mathrm{an}$. As a consequence of the constraint \eqref{eq:transverse deviation}, the spin-wave field will oscillate in the transverse plane $yz$.
Under these assumptions, one has from \eqref{eq:demagnetizing tensor fourier}-\eqref{eq:demag fact Fourier}:
\begin{equation}
    \overleftrightarrow{N}(\bm k)=\overleftrightarrow{N}(k)=\left(\begin{array}{ccc}
       1-s_k  & 0 & 0 \\
        0 &  0  &  0 \\
        0 &  0  & s_k 
    \end{array}\right) \,. \label{eq:N(k) tensor along x}
\end{equation}

The dispersion relation associated with BV inertial spin waves can be found by setting $\alpha=0$ and solving Eq.~\eqref{eq:det inv_Chi zero SW}. To this end, we define:
\begin{align}
    \omega_{11}&=h_0+\ell_\mathrm{ex}^2k^2 \,,\\
    \omega_{22}&=h_0+s_k+\ell_\mathrm{ex}^2k^2 \,.
\end{align}
The perturbation analysis developed in Appendix provides the roots $\pm\omega_N,\pm\omega_P$ in the form \eqref{eq:gen pert omega_N},\eqref{eq:gen pert omega_P}.
For $\xi\ll 1$ and moderately small wavenumber $|k|$, one has from Eqs.~\eqref{eq:gen pert omega_N small xi},\eqref{eq:gen pert omega_P small xi}:
\begin{align}
    \omega_N(k)&\overset{\xi\ll 1}{\approx}\frac{1}{\xi}+\left(k^2\,{\ell_\mathrm{ex}}^2+h_{0}+\frac{s_{k}}{2}\right) \,,\label{eq:BV omega_N small xi}\\   
    \omega_P(k)&\overset{\xi\ll 1}{\approx}
    \sqrt{\left(k^2\,{\ell_\mathrm{ex}}^2+h_{0}\right)\left(k^2\,{\ell_\mathrm{ex}}^2+h_{0}+s_{k}\right)} \cdot \\
&\left(1-\xi \left(k^2\,{\ell_\mathrm{ex}}^2+h_{0}+\frac{s_{k}}{2}\right)\right) \,. \label{eq:BV omega_FMR small xi}
\end{align}
Equation \eqref{eq:BV omega_N small xi} immediately shows the baseline resonant frequency $1/\xi$ in the THz range plus the wavevector dependent contribution from exchange ($\sim k^2$) and magnetostatics ($\sim s_k$).
Conversely, Eq.~\eqref{eq:BV omega_FMR small xi} expresses the influence of inertia on the classical backward volume precessional spin waves, which results in a redshift
as it happens for the macrospin case (see formula \eqref{eq:resonance frequencies prec}).

The FWHM linewidths are given by Eqs.~\eqref{eq:gen pert delta_omega_N small xi},\eqref{eq:gen pert delta_omega_P small xi}:
\begin{align}
 2\Delta\omega_N&\overset{\xi,\alpha\ll 1}\approx
\frac{2\alpha}{\xi}
-\alpha(2h_0 + s_k + 2k^2\ell_\mathrm{ex}^2) \\
&+\frac{3}{2}\alpha\xi(2h_0 + s_k + 2k^2\ell_\mathrm{ex}^2)^2  \,, \label{eq:BV delta_omega_N}\\
    2\Delta\omega_P&\overset{\xi,\alpha\ll 1}\approx\alpha(2h_0 + s_k + 2k^2\ell_\mathrm{ex}^2) \\
    &-\frac32\alpha\xi(2h_0 + s_k + 2k^2\ell_\mathrm{ex}^2)^2 \,. \label{eq:BV delta_omega_P}
\end{align}
%

Equations \eqref{eq:BV delta_omega_N} and \eqref{eq:BV delta_omega_P} express the dependence of spectral linewidth on damping, inertia and static internal field.

\subsubsection{Damon-Eschbach (DE) inertial spin waves}

\begin{figure}
    \centering
    \includegraphics[width=7.5cm]{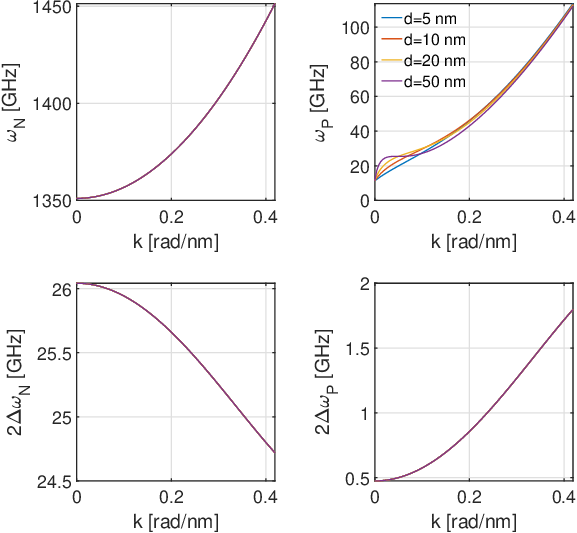}
    \caption{Dispersion relations and FWHM linewidths for nutational and precessional DE inertial spin waves in cobalt thin films as a function of film thickness $d$. Values of parameters same as Fig.~\ref{fig:BV_iSW_dispersion}. Upper panels show nutational and precessional dispersion relations; lower panels show the corresponding FWHM linewidths.}
    \label{fig:DE_iSW_dispersion}
\end{figure}

In the DE configuration, the wavevector $\bm k$ is transverse to the equilibrium $\bm m_0$ and to the static applied field $\bm h_\mathrm{dc}$ which are parallel (see Fig.~\ref{fig:SW geometries flat film}(b)). We assume, for the sake of simplicity, that $\bm k=k\bm e_x, \bm m_0=\bm e_\mathrm{an}=\bm e_y, \bm h_\mathrm{dc}=h_{ay}\bm e_y$, which imply $h_0=h_{ay}+h_\mathrm{an}$. As a consequence of the constraint \eqref{eq:transverse deviation}, the spin-wave field will oscillate in the transverse plane $xz$. The magnetostatic tensor retains the same expression \eqref{eq:N(k) tensor along x}.
We define:
\begin{align}
    \omega_{11}&=h_0+1-s_k+\ell_\mathrm{ex}^2k^2 \,,\\
    \omega_{22}&=h_0+s_k+\ell_\mathrm{ex}^2k^2 \,.
\end{align}
The perturbation analysis developed in Appendix provides the roots $\pm\omega_N,\pm\omega_P$ in the form \eqref{eq:gen pert omega_N},\eqref{eq:gen pert omega_P}.

As in the previous case, assuming $\xi\ll 1$ and moderately small wavenumber $|k|$, 
one has the following simpler expressions:
\begin{align}
    \omega_N(k)&\overset{\xi\ll 1}{\approx}\frac{1}{\xi}+\left(h_0+\frac{1}{2}+\ell_\mathrm{ex}^2k^2\right) \,,\label{eq:DE omega_N small xi}\\
    \omega_P(k)&\overset{\xi\ll 1}{\approx}
    \sqrt{\left(h_0+1-s_k+\ell_\mathrm{ex}^2k^2\right)\left(h_0+s_k+\ell_\mathrm{ex}^2k^2\right)} \cdot \\
&\left(1-\xi \left(h_0+\frac{1}{2}+\ell_\mathrm{ex}^2k^2\right)\right) \,. \label{eq:DE omega_FMR small xi}
\
\end{align}
As discussed for the BV case, Eq.~\eqref{eq:DE omega_N small xi} expresses the dispersion relation for nutational inertial spin waves oscillating at THz frequencies, while Eq.~\eqref{eq:DE omega_FMR small xi} evidences again the redshift due to magnetic inertia on the classical precessional DE spin waves dispersion relation.  Interestingly, the nutational dispersion relation is independent of the wavevector-dependent demagnetizing factor $s_k$, and hence of the film thickness within the adopted approximation. The same occurs for the nutational and precessional FWHM linewidths. Conversely, the precessional DE spin waves dispersion relation exhibits the well-known dependence\cite{Kalinikos1986} on the film thickness.

The FWHM linewidths are given by Eqs.~\eqref{eq:gen pert delta_omega_N small xi},\eqref{eq:gen pert delta_omega_P small xi}:
\begin{align}
 2\Delta\omega_N&\overset{\xi,\alpha\ll 1}\approx
\frac{2\alpha}{\xi}
-\alpha(2h_0 +1 + 2k^2\ell_\mathrm{ex}^2) \\
&+\frac{3}{2}\alpha\xi(2h_0 + 1+ 2k^2\ell_\mathrm{ex}^2)^2  \,, \label{eq:DE delta_omega_N}\\
    2\Delta\omega_P&\overset{\xi,\alpha\ll 1}\approx\alpha(2h_0 +1 + 2k^2\ell_\mathrm{ex}^2) \\
    &-\frac32\alpha\xi(2h_0 + 1 + 2k^2\ell_\mathrm{ex}^2)^2 \,. \label{eq:DE delta_omega_P}
\end{align}
Equations \eqref{eq:DE delta_omega_N} and \eqref{eq:DE delta_omega_P} express the dependence of spectral linewidth on damping, inertia and static internal field.

\subsubsection{Forward volume (FV) inertial spin waves}

\begin{figure}
    \centering
    \includegraphics[width=7.5cm]{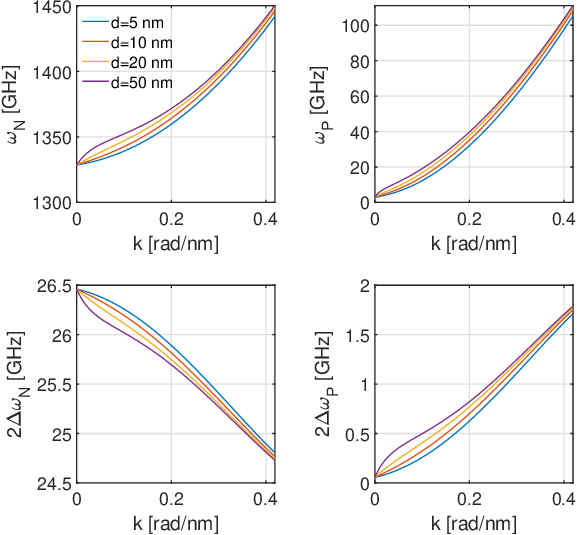}
    \caption{Dispersion relations and FWHM linewidths for nutational and precessional FV inertial spin waves in cobalt thin films as a function of film thickness $d$. Values of parameters same as Fig.~\ref{fig:BV_iSW_dispersion}. Upper panels show nutational and precessional dispersion relations; lower panels show the corresponding FWHM linewidths.}
    \label{fig:FV_iSW_dispersion}
\end{figure}
In the FV configuration, the wavevector $\bm k$ is in-plane while the equilibrium $\bm m_0$ and the static applied field $\bm h_\mathrm{dc}$ are parallel and out-of-plane (see Fig.~\ref{fig:SW geometries flat film}(c)). We assume, for the sake of simplicity, that $\bm k=k\bm e_x, \bm m_0=\bm e_\mathrm{an}=\bm e_z, \bm h_\mathrm{dc}=h_{az}\bm e_z$, which imply $h_0=h_{az}+h_\mathrm{an}$. As a consequence of the constraint \eqref{eq:transverse deviation}, the spin-wave field will oscillate in the transverse plane $xy$. The magnetostatic tensor retains the same expression \eqref{eq:N(k) tensor along x}. We define:
\begin{align}
    \omega_{11}&=h_0+1-s_k+\ell_\mathrm{ex}^2k^2 \,,\\
    \omega_{22}&=h_0+\ell_\mathrm{ex}^2k^2 \,.
\end{align}
The perturbation analysis developed in Appendix
provides the roots $\pm\omega_N,\pm\omega_P$ in the form \eqref{eq:gen pert omega_N},\eqref{eq:gen pert omega_P}.

As in the previous case, assuming $\xi\ll 1$ and moderately small wavenumber $|k|$, 
one has the following simpler expressions:
\begin{align}
    \omega_N(k)&\overset{\xi\ll 1}{\approx}\frac{1}{\xi}+\left(h_0+\frac{1-s_k}{2} +\ell_\mathrm{ex}^2k^2\right) \,,\label{eq:FV omega_N small xi}\\
    \omega_P(k)&\overset{\xi\ll 1}{\approx}
 \sqrt{\left(h_0+1-s_k+\ell_\mathrm{ex}^2k^2\right)\left( h_0+\ell_\mathrm{ex}^2k^2\right)} \\
 &\qquad \cdot \left[ 1 - \xi \left(h_0+\frac{1-s_k}{2} +\ell_\mathrm{ex}^2k^2\right) \right]
    \,. \label{eq:FV omega_FMR small xi}
\end{align}
As discussed for BV and DE cases, Eq.~\eqref{eq:FV omega_N small xi} expresses the dispersion relation for nutational inertial spin waves oscillating at THz frequencies, while Eq.~\eqref{eq:FV omega_FMR small xi} evidences again the redshift due to magnetic inertia on the classical precessional FV spin waves dispersion relation.

The FWHM linewidths are given by Eqs.~\eqref{eq:gen pert delta_omega_N small xi},\eqref{eq:gen pert delta_omega_P small xi}:
\begin{align}
 2\Delta\omega_N&\overset{\xi,\alpha\ll 1}\approx
\frac{2\alpha}{\xi}
-\alpha(2h_0 +1 - s_k + 2k^2\ell_\mathrm{ex}^2) \\
&+\frac{3}{2}\alpha\xi(2h_0 +1 - s_k + 2k^2\ell_\mathrm{ex}^2)^2  \,, \label{eq:FV delta_omega_N}\\
    2\Delta\omega_P&\overset{\xi,\alpha\ll 1}\approx\alpha(2h_0 + 1 - s_k + 2k^2\ell_\mathrm{ex}^2) \\
    &-\frac32\alpha\xi(2h_0 +1 - s_k + 2k^2\ell_\mathrm{ex}^2)^2 \,. \label{eq:FV delta_omega_P}
\end{align}


Equations \eqref{eq:FV delta_omega_N} and \eqref{eq:FV delta_omega_P} express the dependence of spectral linewidth on damping, inertia and static internal field. 

\subsection{Spin waves in curved magnetic nanostrips}

The theory has been recently extended\cite{dAquino2025} to treat the interesting situation of inertial spin waves propagating on twisted ferromagnetic ultrathin nanostrips, which is becoming increasingly attractive owing to recent advances in nanofabrication techniques
for three-dimensional magnetic structures\cite{gubbiotti2025}. Here we report the main analytical findings related to the determination of the dispersion relation that arises from the nontrivial solution of unforced Eq.~\eqref{eq:linearized iLLG fourier} ($\delta\bm h_\mathrm{ac}=0$). 

\begin{figure}[t]
    \centering
    \includegraphics[width=\linewidth]{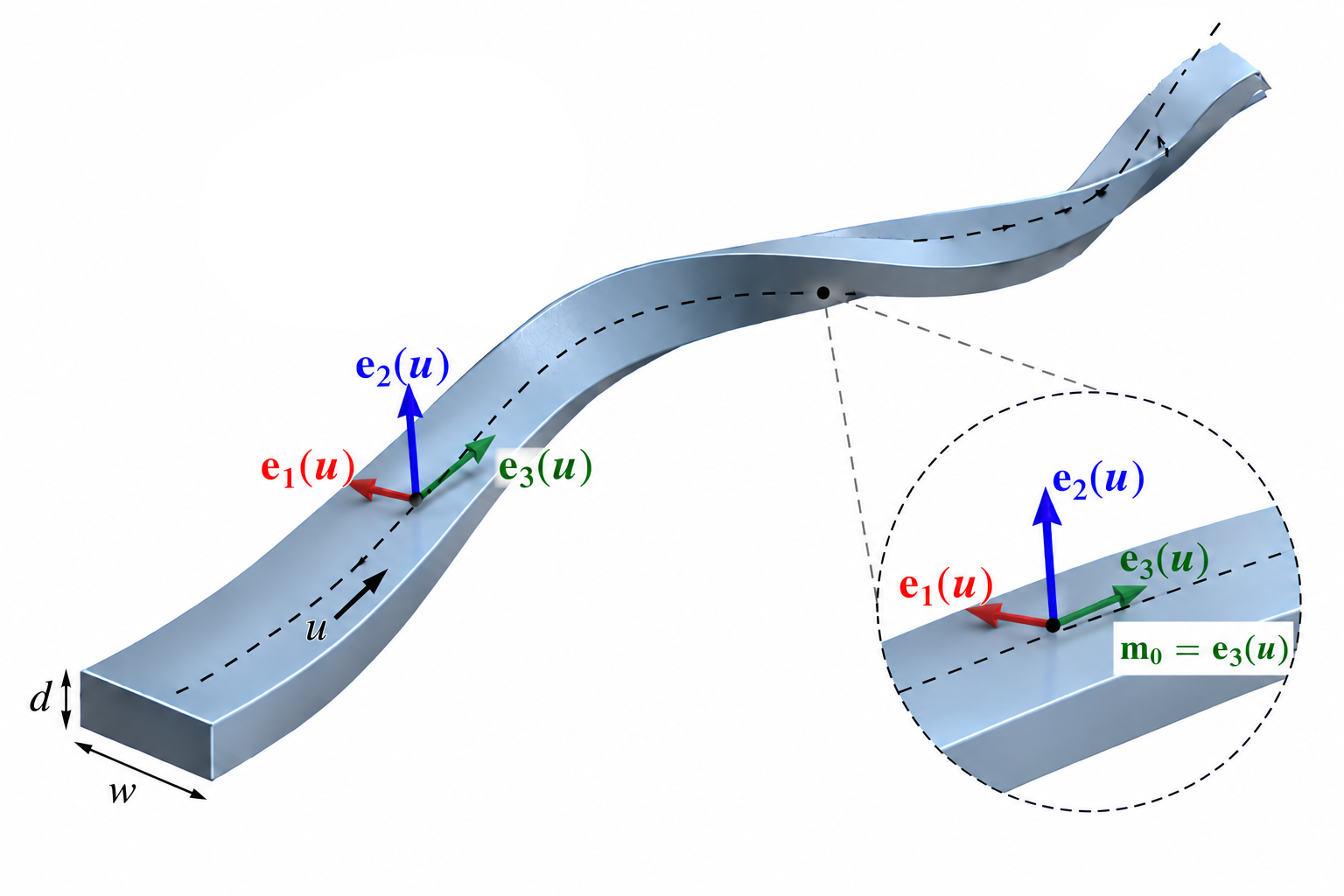}
    \caption{Sketch of a curved ultrathin strip with equilibrium magnetization $\bm m_0$ tangential to the axis of the strip. $(\bm e_1, \bm e_2, \bm e_3)$ is the local reference frame defined along the curvilinear abscissa $u$ induced by the parametric representation of the strip surface.}
    \label{fig:curved strip}
\end{figure}

To this end, a ferromagnetic ultrathin nanostrip, modeled as a ruled surface with thickness $d\sim \ell_\mathrm{ex}$, width $w$ and length $L$ such that $L\gg w$, having axis curve with curvature $\kappa$ and torsion $\tau$ is considered. We assume that the equilibrium magnetization $\bm m_0$ is tangential to the axis of the strip (see Fig.~\ref{fig:curved strip}). 

Following the same approach developed in appendix \ref{sec:appendix}, we introduce the following quantities:
\begin{align}
    \omega_{11}(k)&=h_0 + N_1 + \ell_{\mathrm{ex}}^2(k^2 + \tau^2 + \kappa^2) \,,\\
    \omega_{22}(k)&=h_0 + N_2 + \ell_{\mathrm{ex}}^2(k^2 + \tau^2) \,,
\end{align}
where $N_1$ and $N_2$ are the demagnetizing factors
along the transverse in-plane and transverse out-of-plane
directions, respectively.
The resulting analytical dispersion relations, obtained in
Ref.~\cite{dAquino2025}, are accurate
in the range $|k\ell_\mathrm{ex}/\pi|\leq 1$,
which extends to the large wavenumbers associated
with ultrashort spin waves.

Specifically, assuming that $\xi \ll 1$, one has the following expressions:
\begin{align}
\omega_N
&\approx
\frac{1}{\xi}
+h_0+\frac{N_1+N_2}{2}+\ell_{\mathrm{ex}}^2\left(k^2 + \tau^2 + \frac{\kappa^2}{2}+2k\tau\right)
\,, \label{eq:dispersion relation nutation} \\
    \omega_P&\approx 
\frac{\sqrt{\omega_{11}\,\omega_{22}}}{\sqrt{1+\xi(\omega_{11}+\omega_{22}) }}-2k\tau\ell_\mathrm{ex}^2 \label{eq:dispersion relation precession}
\end{align}
which show that the oscillations occur in the THz range ($\xi \sim 4.5\times 10^{-2}$ in permalloy \cite{Neeraj2021inertial} means $\gamma M_s/(2\pi\xi)\sim 630$ GHz).
Equations \eqref{eq:dispersion relation nutation} and \eqref{eq:dispersion relation precession} indicate that both nutational and precessional resonances are influenced by the strip geometry, specifically by its curvature $\kappa$ and torsion $\tau$, which enter through $\omega_{11}(k)$ and $\omega_{22}(k)$, generally leading to an upward shift of the fundamental ($k=0$)  frequencies. In addition, from Eq.~\eqref{eq:dispersion relation precession} it follows that inertial effects can only produce a reduction of the frequency, i.e., a redshift, when compared to the classical non-inertial limit $\xi=0$.

Furthermore, in the case of a twisted geometry with $\tau \neq 0$, both the precessional and nutational dispersion branches become asymmetric, revealing the nonreciprocal nature of spin-wave propagation due to the appearance of a geometric (Berry) phase accumulation\cite{dAquino2025}. In contrast, curvature $\kappa$ merely shifts the frequencies without breaking the symmetry of the dispersion relation, thus preserving reciprocity. Inspecting the terms proportional to $\tau$ in Eqs.~\eqref{eq:dispersion relation nutation}--\eqref{eq:dispersion relation precession}, one finds that torsion induces the same type of asymmetry in both branches, but with opposite signs for precessional and nutational modes. This distinction may serve as a useful signature for identifying inertial effects in experimental observations. Moreover, it has been shown\cite{dAquino2025} that FWHM linewidths $2\Delta\omega_N,2\Delta\omega_P$ exhibit the same asymmetry coming from nonzero torsion as it appears in the dispersion relations, and that $\Delta\omega_N$ (resp. $\Delta\omega_P$) has a maximum (resp. minimum) close to $k=0$. 

The spin-wave wavenumber quantization associated with
nanostrip confinement and nontrivial topology is also
discussed in Ref.~\cite{dAquino2025}, disentangling the intriguing interplay between chirality and topology.

\subsection{Group velocities and inertial spin-wave speed limit}

In all the situations described above, the analytical dispersion relations allow the spin-wave
group velocity to be straightforwardly computed as:
\begin{equation}
    v_g(k)=\gamma M_s \pdv{\omega}{k} \,.\label{eq:group velocity}
\end{equation} 
It is apparent that, for flat thin films where the dependence on wavenumber in Eqs.~\eqref{eq:BV omega_N small xi},\eqref{eq:BV omega_FMR small xi},\eqref{eq:DE omega_N small xi},\eqref{eq:DE omega_FMR small xi},\eqref{eq:FV omega_N small xi},\eqref{eq:FV omega_FMR small xi} occurs through the amplitude $\abs{\bm k}$ in the magnetostatics-related terms $s_k$ and through $k^2$ in exchange-related terms, both precessional and nutational spin waves clearly exhibit reciprocal propagation. Conversely, the geometrical twist present in curved strips gives rise to asymmetry (with opposite signs for precessional and nutational spin waves) due to the term $\pm 2k\tau\ell_\mathrm{ex}^2$ in Eqs.~\eqref{eq:dispersion relation nutation},\eqref{eq:dispersion relation precession}, respectively. 

Finally, for all the considered situations of inertial spin waves propagating in ultrathin films and strips, a general result holds for the group velocity in the large wavenumber limit due to the hyperbolic mathematical nature of the iLLG Eq.~\eqref{eq:iLLG equation} that is profoundly different from the classical precessional LLG dynamics. Specifically, for large $|\bm k|$, one can compute the dispersion relation retaining in Eq.~\eqref{eq:det inv_Chi zero SW} with $\alpha=0$ only terms in $\omega$ and $k^2$ and neglecting the remaining $\mathcal{O}(1)$ terms (i.e. static internal field, anisotropy and magnetostatic field), which leads to:
\begin{equation}
    \det\{\mathcal{P}_\perp[i\omega \Lambda(\bm m_0) - (\xi \omega^2 - l_{ex}^2 k^2) \mathcal{I}]\}=0 \,. \label{eq:det inv_Chi large k}
\end{equation}
According to Eq.~\eqref{eq:gen characteristic poly sym}, the computation of the determinant in Eq.~\eqref{eq:det inv_Chi large k} yields
\begin{equation}
(\xi \omega^2 - l_{ex}^2 k^2)^2 - \omega^2=0 \,,
\end{equation}
which in turn produces the following dispersion relation
\begin{equation}\label{eq:dispersion relation large k}
\omega(k) = \frac{\pm 1 \pm \sqrt{1 + 4\xi \ell_{ex}^2 k^2}}{2\xi} \,,
\end{equation}
where the pair of $\pm$ determines the four independent solutions.
It is apparent that, for large wavenumber such that $4\xi \ell_\mathrm{ex}^2k^2\gg 1$, from Eq.~\eqref{eq:dispersion relation large k} one ends up with the asymptotic limits for the spin-wave frequencies:
\begin{equation}
    \omega(k)\overset{4\xi \ell_\mathrm{ex}^2k^2\gg 1}\approx \pm \frac{1}{2\xi}\pm\frac{\ell_\mathrm{ex}}{\sqrt{\xi}}\,|k| \,,
\end{equation}
which according to Eq.~\eqref{eq:group velocity} implies the ultimate speed limit for inertial spin waves:
\begin{equation}\label{eq:iSW speed limit}
    v_g(k\rightarrow\pm \infty)=\pm \frac{\ell_\mathrm{ex}}{t_\mathrm{in}} \,.
\end{equation}
The ultimate speed limit \eqref{eq:iSW speed limit}, for usual ferromagnetic materials, is of the order of several thousand m/s (e.g. for cobalt $t_\mathrm{in}=0.653\,$ps, $\mu_0M_s=1.6~$T, $A_\mathrm{ex}=13\,$pJ/m, $v_g(k\rightarrow\infty)\approx 5470\,$m/s). 

Interestingly, this limiting velocity is of the same order
of magnitude as the propagation velocities of surface acoustic
waves (SAWs), typically of the order of a few $10^3$ m/s in
solid substrates, which are known to couple efficiently to
conventional spin waves through magnetoelastic interactions
\cite{Geilen2022}.

\section{Micromagnetic simulation of inertial magnetization dynamics}\label{sec:numerical}

\subsection{Overview of literature}

The numerical simulation of inertial magnetization dynamics presents
additional challenges compared with conventional micromagnetics.
Besides the very short time scales associated with nutational dynamics,
the presence of the second-order time derivative in the inertial
Landau--Lifshitz--Gilbert (iLLG) equation effectively enlarges the
dynamical phase space and requires numerical schemes capable of
accurately resolving both the conventional precessional and the
ultrafast nutational motion. Furthermore, reliable discretizations
should reproduce, as closely as possible, the fundamental geometrical
and energetic properties of the continuous dynamics, including the
unit-length constraint of magnetization and the generalized
free-energy balance discussed in Sec.~III.

Some of the numerical tools used for inertial micromagnetics originate
from methods previously developed for the conventional LLG equation.
In particular, geometric integration based on the implicit midpoint
rule was introduced for micromagnetic dynamics by d'Aquino et al.~\cite{dAquino2005}. Such an approach preserves the
magnetization-length constraint independently of the time step and
provides favorable stability properties for the integration of the
strongly nonlinear LLG dynamics. Efficient quasi-Newton strategies
were also developed to deal with the large nonlinear algebraic systems
arising from the implicit discretization. In parallel, numerical
techniques for the linearized LLG equation were developed to investigate
the frequency-domain response of spatially nonuniform micromagnetic
systems. An operator formulation for the computation of magnetization
eigenmodes in complex micromagnetic systems was proposed in Ref.~\cite{daquino_novel_2009}, while a large-scale frequency-domain
approach for directly computing the magnetization response to
time-harmonic excitation was subsequently developed in Ref.~\cite{daquino_micromagnetic_2023}. These techniques exploit
matrix-free evaluations of the micromagnetic effective-field operator
and iterative Krylov-subspace methods, thus avoiding the explicit
assembly of the dense matrices associated with long-range
magnetostatic interactions.

Numerical methods specifically designed for magnetization dynamics
with inertial effects started to appear together with the development
of the iLLG model. An early contribution was given by Moumni and
Tilioua~\cite{Moumni201695}, who proposed a semi-implicit
finite-difference scheme for a model of magnetization dynamics
including an inertial acceleration term. The authors investigated the
stability of the discretization and performed numerical experiments
illustrating the influence of inertia on the short-time magnetization
response.

A significant step toward a rigorous numerical treatment of the
spatially inhomogeneous iLLG equation was made by
Ruggeri~\cite{Ruggeri20221199}, who proposed and analyzed two
fully discrete finite-element schemes. One formulation is based on
magnetization and its velocity, whereas the other exploits an
equivalent first-order formulation involving magnetization and an
angular-momentum variable. The schemes preserve the unit-length
constraint at the vertices of the computational mesh, and convergence
of suitable subsequences of the discrete solutions toward weak
solutions of the iLLG equation was established.

More recently, Aldé et al.~\cite{alde2026bdf2} proposed and analyzed a fully discrete finite-element scheme for iLLG dynamics based on a second-order backward differentiation formula (BDF2), with a suitable treatment of the inertial term. The authors established well-posedness, unconditional stability and convergence towards weak solutions, together with a discrete energy law and error estimates for the magnetization unit-length constraint.

Another numerical approach was developed by
 Li et al.~\cite{Li2022182}, who proposed a second-order
semi-implicit finite-difference method for the iLLG equation. The
scheme combines centered approximations of the inertial term with a
suitable treatment of the nonlinear contributions, resulting in
linear algebraic systems at each time step. Its solvability and
accuracy were analyzed, and numerical experiments were used to
illustrate both the short-time inertial regime, characterized by
magnetization nutation, and the transition toward the conventional
longer-time precessional dynamics.

More recently, Chen et al.~\cite{Chen2024} proposed an implicit
central finite-difference discretization of the iLLG equation,
together with an iterative procedure for solving the resulting
nonlinear algebraic problem. A convergence analysis was developed,
establishing second-order accuracy in space and time under the
assumptions considered therein. The numerical scheme was also analyzed
with respect to the characteristic structural properties of the iLLG
dynamics and applied to the simulation of short-time inertial
magnetization processes.

A complementary approach particularly suited to large-scale
micromagnetic simulations was developed in Ref.~\cite{daquino2024midpoint}. Two second-order time-stepping
techniques based on the implicit midpoint rule were proposed. The
first one applies the midpoint discretization to an equivalent
first-order dynamical system involving magnetization and an additional
angular-momentum variable. This formulation exactly reproduces at the
discrete level important qualitative properties of the continuous
iLLG dynamics, including magnetization-length conservation and, in
the appropriate cases, angular-momentum and generalized-energy
conservation or dissipation. The second technique combines the
implicit midpoint rule with a multistep approximation of the inertial
term, allowing the numerical evolution to be expressed directly in
terms of magnetization. This substantially reduces the number of
unknowns and facilitates the incorporation of inertial effects into
existing micromagnetic solvers. Large-scale quasi-Newton techniques
were introduced in both cases to efficiently treat the nonlinear
systems resulting from the implicit discretization.

Besides the development and analysis of numerical schemes, spatially
resolved micromagnetic simulations have played an important role in
revealing physical effects that cannot be captured within the
macrospin approximation. In particular, analytical spin-wave theory was combined
with frequency- and time-domain micromagnetic simulations
in Ref.~\cite{daquino2023micromagnetic} to investigate
inertial spin waves in confined ferromagnetic nanodots.
The simulations demonstrated that THz excitation can generate
nutational spin waves with wavelengths of the order of tens of
nanometers and allowed their finite propagation velocity to be
directly observed in the time domain. This provided a numerical
connection between the inertial spin-wave dispersion relations
discussed in Sec.~IV and the spatial magnetization patterns expected
in realistic confined structures.

Micromagnetic simulations are also particularly important in the
strongly nonlinear regime. Inertial magnetization switching driven by
intense ultrashort magnetic-field pulses was investigated by Neeraj
et al.~\cite{neeraj2022inertial}, showing how magnetic inertia can
substantially modify the switching trajectories on picosecond time
scales. More recently, Fortunati et al.~\cite{Fortunati_2026}
developed an analytical description of ultrafast precessional
switching in the inertial regime and compared it with both macrospin
and full micromagnetic simulations. A more recent investigation \cite{fortunati2026controlled} has proposed a controlled switching strategy based on exponentially decaying transverse magnetic fields, combining approximate analytical modelling, energy-based conditions for preventing backhopping, and spatially resolved micromagnetic simulations to assess the robustness of magnetization reversal.
Such comparisons are especially
useful for assessing when spatially coherent approximations remain
valid and when exchange and magnetostatic interactions generate
significant spatial nonuniformities.

In the following, we review numerical techniques for the simulation
of iLLG dynamics with particular emphasis on methods suitable for
large-scale micromagnetic computations. We first consider the
frequency-domain formulation of the linearized problem, which provides
an efficient route to the calculation of the steady-state response and
the excitation of inertial spin waves. Then, among the available
numerical approaches for time-domain discretization, we focus on the
geometric midpoint schemes introduced in
Ref.~\cite{daquino2024midpoint}, discussing their conservation
properties and computational implementation. These methods provide
a particularly convenient framework for preserving the qualitative
properties of iLLG dynamics and for extending existing large-scale
micromagnetic simulation codes. Finally, representative micromagnetic
simulations of ultrashort inertial spin-wave propagation and
ultrafast magnetization switching are presented.


To this end, we proceed to the numerical discretization of the iLLG equation. 
 In the following, we will refer to spatially semi-discretized equations on a collection of $N$ mesh points $(\bm r_j)_{j=1}^N$ associated with the computational cells of volume $V_j$. This description is quite general and works both for finite-difference and finite-element methods. In what follows, the dot product denotes the appropriate volume-weighted discrete inner product; for a uniform finite-difference mesh it differs from the Euclidean inner product only by a common constant factor.
 
 We will denote as $\underline{\bm m}(t)=(\bm m_1,\ldots , \bm m_N)^T$, $\underline{\bm w}(t)=(\bm w_1,\ldots, \bm w_N)^T$$ \in \mathbb{R}^{3N}$ (the notation $^T$ means matrix transpose) the mesh vectors containing all cell vectors $\bm m_j(t),\bm w_j(t) \in \mathbb{R}^3$ with $j=1,\ldots,N$.

Moreover, we will use the operator notation for the cross-product for both cell and mesh vectors, namely:
\begin{equation}
    \Lambda({\bm v})\cdot \bm w=\bm v\times \bm w\,,\quad \underline{\Lambda}({\underline{\bm v}})\cdot \underline{\bm w}=(\bm v_1\times \bm w_1, \ldots, \bm v_N\times \bm w_N)^T \,,
\end{equation}
meaning that the latter operator is a skew-symmetric $3N\times 3N$ block-diagonal operator that provides cross product of homologous cell vectors.

\subsection{Frequency-domain simulation of iLLG dynamics}\label{sec:frequency domain mumag}

The starting point is the linearized iLLG equation \eqref{eq:linearized iLLG} rewritten in the time-harmonic phasor domain:
\begin{equation}\label{eq:linearized iLLG frequency domain}
    i\omega \delta\tilde{\bm m} = \bm m_0\times \left(-\delta\tilde{\bm h}_\mathrm{eff}+h_0 \delta\tilde{\bm m} + i\omega\alpha\delta\tilde{\bm m}-\xi\omega^2\delta\tilde{\bm m}-\delta\tilde{\bm h}_\mathrm{ac}\right) \,,
\end{equation}
where the phasor notation $\tilde{\bm v}$ has been used for vector fields such that $\bm v(\bm r,t)=\Re[\tilde{\bm v}(\bm r)e^{i\omega t}]$. By using the cross-product operator $\Lambda$, one obtains:
\begin{align}
    \delta\tilde{\bm m}&=\chi(\omega)\delta\tilde{\bm h}_\mathrm{ac}=\Big\{\mathcal{P}_\perp[i\omega\Lambda(\bm m_0)+\mathcal{C}  \\
    &+(h_0-\xi\omega^2+i\omega\alpha)\mathcal{I}]  \Big \}^{-1}\!\!\!\! \cdot \delta\tilde{\bm h}_\mathrm{ac} \,. \label{eq:iLLG mumag freq resp}
\end{align}
The latter equation can be interpreted as a linear system
\begin{equation}\label{eq:linear forced response operator form}
    \mathcal{L}\delta\tilde{\bm m}=\delta\tilde{\bm h}_\mathrm{ac} \,,
\end{equation}
where $\mathcal{L}=\mathcal{P}_\perp[i\omega\Lambda(\bm m_0)+\mathcal{C}+(h_0-\xi\omega^2+i\omega\alpha)\mathcal{I}]$ is the linear operator to be inverted in order to determine the unknown $\delta\tilde{\bm m}$.
This can be efficiently addressed by using large-scale techniques for the computation of the effective field operator $\delta\tilde{\bm h}_\mathrm{eff}(\tilde{\bm v})=-\mathcal{C}\tilde{\bm v}$ as described in Ref.~\cite{daquino_micromagnetic_2023}. Here we briefly summarize the main steps.

In fact, one can consider the discretized version of Eq.~\eqref{eq:linear forced response operator form}:
\begin{equation}\label{eq:linear forced response discrete operator form}
    \underline{L}\cdot \delta\tilde{\underline{\bm m}}=\delta\tilde{\underline{\bm h}}_\mathrm{ac} \,,
\end{equation}
where the matrix $\underline{L}$ is given by:
\begin{equation}\label{eq:discretized operator L}
    \underline{L}=\underline{P}_{\bm m_0}\cdot \left[i\omega\underline{\Lambda}(\underline{\bm m}_0)+\underline{C}+\underline{H}_0+(-\xi\omega^2+i\omega\alpha)\underline{I}\right]
\end{equation}
where $\underline{P}_{\bm m_0}=\text{diag}[I-\bm m_{01}\otimes\bm m_{01}, \ldots ,I-\bm m_{0N}\otimes\bm m_{0N}]$ and $\underline{H}_0=\text{diag}[h_{01}I, \ldots ,h_{0N}I]$ ($\underline{I}$ and $I$ are the $3N\times 3N$ and $3\times 3$ identity matrices, respectively).

We observe that the matrix $\underline{L}$ given by Eq.~\eqref{eq:discretized operator L} in the linear system \eqref{eq:linear forced response discrete operator form} is full owing to the magnetostatic (all-to-all site interaction) operator, while the other terms are sparse, being either diagonal or $3\times 3$ block-diagonal. To overcome such a bottleneck, one can use Krylov-subspace methods such as GMRES, which only requires the evaluation of the action of the operator $\underline{L}$ on a vector $\underline{\bm v}$, namely the matrix-vector product $\underline{L}\cdot\underline{\bm v}$. Thus, one can separate the evaluation of the effective-field term
from the remaining sparse contributions, leveraging large-scale methods (e.g. Fast Fourier Transform, hierarchical matrices, etc.) for the computation of $\underline{C}\cdot\underline{\bm v}=-\delta\underline{\bm h}_\mathrm{eff}(\underline{\bm v})$, which can reduce the computational cost to $O(N \log N)$ or even $O(N)$.  


\begin{figure*}[t]
    \centering
    \includegraphics[width=15cm]{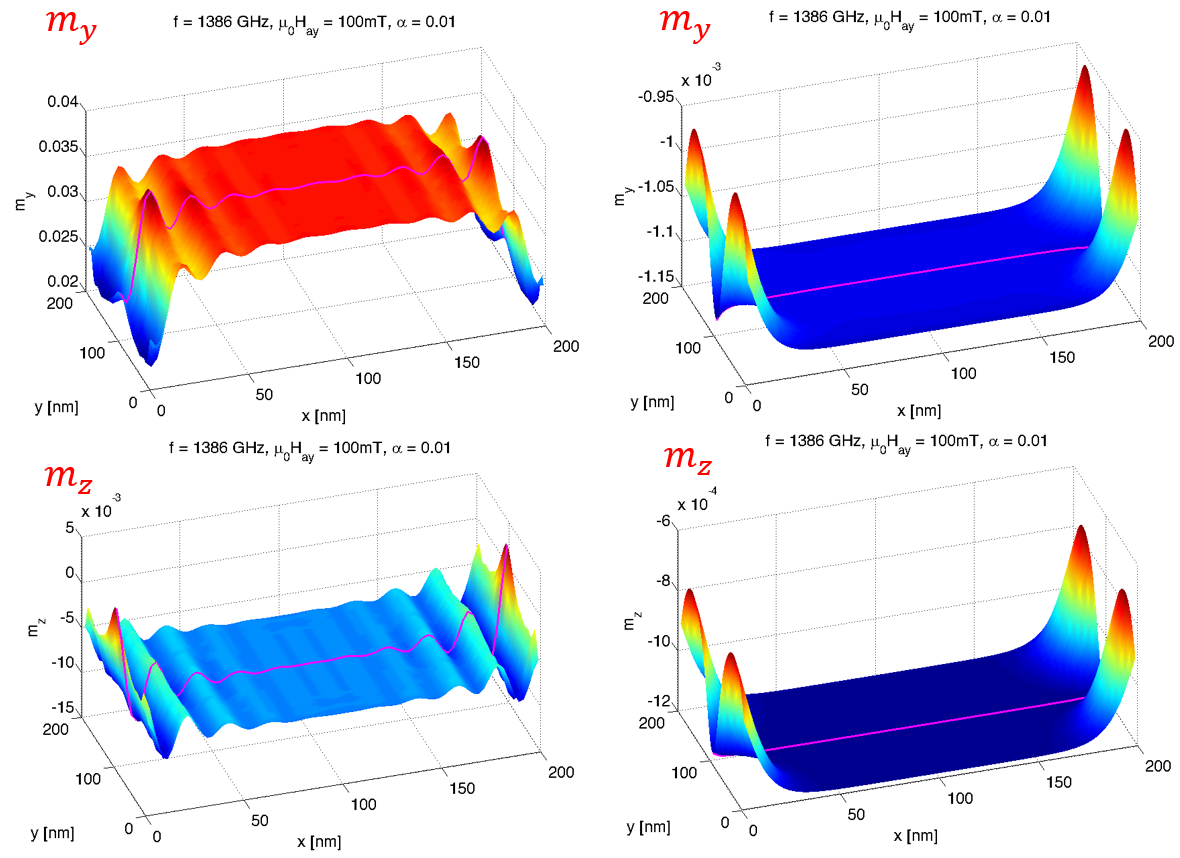}
    \caption{Snapshot at $t = 0$ of the steady-state ac magnetization-response components $m_y$ and $m_z$ at driving frequency $f = 1386$~GHz. The color code represents the value of the corresponding magnetization component at each spatial location ranging from minimum (blue) to maximum (red). The left panels refer to inertial LLG dynamics whereas the right panels refer to classical LLG dynamics. Magenta solid lines refer to $m_y(x),m_z(x)$ sampled in the middle of the nanodot at $y = 100\,$nm.}
    \label{fig:frequency domain micromagnetic simulation}
\end{figure*}

The aforementioned frequency-domain technique has  been successfully used to assess the excitation of ultrashort wavelength inertial spin waves~\cite{daquino2023micromagnetic} in a square $200\times 200\times 5$ nm$^3$ cobalt thin-film nanodot, with the following material parameters: $\gamma=2.211\times 10^5$ m$\,$A$^{-1}\,$s$^{-1}$, $\mu_0 M_s=1.6$ T, $A=13$ pJ/m ($l_\mathrm{ex}=3.57$ nm), $t_\mathrm{in}=0.653$ ps $(\xi=0.0338)$. 
The nanodot is saturated along the $x$ axis by a static field $\mu_0 H_{ax}=100$ mT. 
We consider the ac steady-state magnetization dynamics
of the nanodot driven by a spatially uniform,
linearly polarized ac field transverse to the equilibrium magnetization with frequency $f=1386$ GHz and amplitude $\mu_0 H_{ay}=100$ mT.  
The magnetization is initially in the remanent equilibrium configuration under the static external magnetic field, which is mainly oriented along the $x$ axis with deviations located close to the thin film edges parallel to the $y$ axis.


The ac forced magnetization dynamics can be conveniently studied by solving the linearized iLLG in the frequency domain and determining the frequency response and the power spectrum.  The numerical calculations are performed using a
large-scale finite-difference frequency-domain
micromagnetic solver~\cite{daquino_micromagnetic_2023} based on suitable operator formalism~\cite{daquino_novel_2009}, appropriately extended in order to include inertial effects and implemented in the numerical code MaGICo~\cite{MaGICo}. The output of the code is the steady-state ac magnetization response $\delta\hat{\bm m}(\bm r)$ (such that $\delta\bm m(\bm r,t)=\Re[\delta\hat{\bm m}(\bm r) e^{i\omega t}]$) computed for desired values of the external ac field frequency $\omega$ (see Ref.~\cite{daquino_micromagnetic_2023} for further details). 

The result of micromagnetic simulations with $\alpha=0.01$ is reported in Fig.~\ref{fig:frequency domain micromagnetic simulation}. One can clearly see (left panels of Fig.~\ref{fig:frequency domain micromagnetic simulation}) that inertial spin wave dynamics driven at frequency $f>f_N$ involves ultrashort nutation spin waves with wavelength around 20 nm (one can count roughly ten oscillation periods along the $x$ direction), which is in excellent agreement with the nutational branch of the dispersion relation \eqref{eq:BV omega_N small xi} for BV inertial spin waves (see Fig.~\ref{fig:BV_iSW_dispersion}).  On the other hand, when inertia is not considered (right panels of Fig.~\ref{fig:frequency domain micromagnetic simulation}), the oscillation amplitude is two orders of magnitude smaller than in the inertial case, and short-wavelength spin waves are no longer excited.
The time-evolution of the ac steady-state magnetization $\delta\bm m(\bm r,t)=\Re[\delta\tilde{\bm m}(\bm r) e^{i\omega t}]$ can be interpreted as the superposition of two contributions: an almost spatially uniform nutational response and a spatially nonuniform component associated with inertial spin waves of wavelength approximately $20$~nm. The much larger response amplitude in the inertial case is explained by the proximity of the driving frequency $f=1386$~GHz to the nutational resonance $f_N\approx1328$~GHz, which has no counterpart in the classical LLG dynamics. Apart from this amplitude difference, the long-wavelength magnetization profiles remain qualitatively similar.

\subsection{Time-domain simulation of iLLG dynamics}

Time-domain micromagnetic simulations including inertial effects require particular attention due to the doubling of the number of degrees of freedom associated with the vector field $\bm w$. 
In this respect, by using the notations for cell and mesh vectors introduced before, the semi-discretized iLLG equation will read as:
\begin{align}
    \frac{d \underline{\bm m}}{d t}&=\underline{\Lambda}(\underline{\bm w})\cdot \underline{\bm m} = \underline{\bm f}_{\bm m}(\underline{\bm m},\underline{\bm w})\quad, \label{eq:semidiscrete iLLG_m} \\
    \xi\frac{d  \underline{\bm w}}{d t}&=-\underline{\Lambda}(\underline{\bm m})\cdot \underline{\bm w} -\alpha \underline{\bm w} +\underline{\Lambda}({\underline{\bm m}})\cdot \underline{\bm h}_\mathrm{eff} =\underline{f}_{\bm w}(\underline{\bm m},\underline{\bm w})\quad, \label{eq:semidiscrete iLLG_w}
\end{align}
where the discrete effective field $\underline{\bm h}_\mathrm{eff}$ is given by:
\begin{equation}\label{eq:discrete effective field}
    \underline{\bm h}_\mathrm{eff}(\underline{\bm m},t)=-\frac{\partial \underline{g}}{\partial\underline{\bm m}}=-C\cdot \underline{\bm m}(t) + \underline{\bm h}_a(t)\,,
\end{equation}
the matrix $\underline{C}$, self-adjoint with respect to the adopted discrete inner product, plays the role of the effective field operator $\mathcal{C}$ in Eq.~\eqref{eq:effective field operator form} and $\underline{g}(\underline{\bm m},\underline{\bm h}_a)$ is the discrete counterpart of the free energy defined by Eq.~ \eqref{eq:micromagnetic free energy}:
\begin{equation}\label{eq:discrete free energy}
    \underline{g}(\underline{\bm m},\underline{\bm h}_a)= \frac{1}{2} \underline{\bm m}^T\cdot \underline{C} \cdot \underline{\bm m} -\underline{\bm h}_a^T \cdot \underline{\bm m}\,.
\end{equation}
By using the same line of reasoning as for the continuous iLLG equation \eqref{eq:dm_dt}-\eqref{eq:dw_dt} and defining $\underline{\tilde{g}}(\underline{\bm m},\underline{\bm w}, \underline{\bm h_a})=\underline{{g}}(\underline{\bm m},\underline{\bm h_a})+\frac{\xi}{2}|\underline{w}|^2$ as the discrete 'total' energy corresponding to $\tilde{g}$ in the continuous iLLG dynamics (see Eq.~\eqref{eq:generalized free energy}) one can prove\cite{daquino2024midpoint} that the semi-discretized dynamics fulfill conservation properties perfectly analogous to \textbf{Properties \#1,\#2,\#3}:
\begin{align}
 &1)\quad |\bm m_j(t)| = |\bm m_j(t_0)| \, \forall t\geq t_0, \,  j=1,\ldots,N \,, \label{eq:semidiscrete amplitude conservation} \\
&2)\quad ({\bm w_j(t)}\cdot{\bm m_j(t)})=({\bm w_j(t_0)}\cdot{\bm m_j(t_0)})e^{-\frac{\alpha}{\xi} (t-t_0)} \, \\ &\qquad \forall t\geq t_0, \quad  j=1,\ldots,N \,, 
\label{eq:semidiscrete exponential wdotm} \\
&3)\quad  \frac{d}{dt}\Bigg( \underline{g}(\underline{\bm m}(t), \underline{\bm h}_a(t)) + \frac{\xi}{2}\left|\frac{d\underline{\bm m}}{dt}\right|^2 \Bigg)= \\ &\qquad=\frac{d}{dt}\left( \underline{g}(\underline{\bm m}(t), \underline{\bm h}_a(t)) + \frac{\xi}{2}\left|\underline{\bm w} \right|^2 \right)=\frac{d \underline{\tilde{g}}}{dt} = \\ &\qquad=-\alpha \left|\frac{d\underline{\bm m}}{dt}\right|^2 - \underline{\bm m}\cdot \frac{d\underline{\bm h}_a}{dt} \,,\label{eq:semidiscrete energy balance} 
\end{align}
where the assumption of compatible data for $\underline{\bm w}$ and $d\underline{\bm m}/dt$ at the initial time $t_0$ according to Eq.~\eqref{eq:w definition}  is understood.
 
Equations \eqref{eq:semidiscrete iLLG_m}-\eqref{eq:semidiscrete iLLG_w} can be time-discretized as usual by setting
\begin{equation}
    \dv{\underline{\bm m}}{t} \approx \frac{\underline{\bm m}^{n+1}-\underline{\bm m}^n}{\Delta t} \,,\, \dv{\underline{\bm w}}{t} \approx \frac{\underline{\bm w}^{n+1}-\underline{\bm w}^n}{\Delta t} \,,
\end{equation}
in the left-hand side and choosing appropriate expressions for $\underline{\bm m}, \underline{\bm w}$ at the right-hand sides as functions of the unknowns evaluated at time $t^{n+1}$ (implicit methods) or previous time instants (explicit methods). It is well-known that the choice of implicit methods has several advantages from the viewpoint of numerical stability\cite{dAquino2005} given the strongly nonlinear nature of the dynamical system \eqref{eq:semidiscrete iLLG_m}-\eqref{eq:semidiscrete iLLG_w}, but for the iLLG dynamics it leads to a nonlinear system of $6N$ coupled scalar equations in the $6N$ scalar unknowns $\underline{\bm m}^{n+1}, \underline{\bm w}^{n+1}$, which must be solved at each time step. 

Moreover, the numerical solution of Eqs.~\eqref{eq:semidiscrete iLLG_m}-\eqref{eq:semidiscrete iLLG_w} with classical time-stepping techniques will generally not preserve the conservation properties of semi-discretized inertial magnetization dynamics. Thus, such properties will be fulfilled with an accuracy depending on the amplitude of the time-step $\Delta t$. For the classical purely precessional LLG equation, it has been shown\cite{dAquino2005}  that the implicit midpoint rule technique preserves the properties of discrete magnetization dynamics regardless of the time-step. 

In the following, we review the methods proposed in Ref.~\cite{daquino2024midpoint}, which provide large-scale techniques to allow full micromagnetic simulations in the inertial regime at a reasonable computational cost. 

\subsection{Midpoint time-steppings for iLLG dynamics}


The first method, hereafter referred to as implicit midpoint rule (IMR), is based on discretization of Eqs.~\eqref{eq:semidiscrete iLLG_m}-\eqref{eq:semidiscrete iLLG_w} at time $t^{n+\frac{1}{2}}=t^n+\Delta t/2$ with the following second-order midpoint formulas:
\begin{equation}\label{eq:mp formulas}
    \underline{\bm m}^{n+\frac{1}{2}}= \frac{\underline{\bm m}^{n+1}+ \underline{\bm m}^{n}}{2} \,\quad \underline{\bm w}^{n+\frac{1}{2}}= \frac{\underline{\bm w}^{n+1}+ \underline{\bm w}^{n}}{2} \,,
\end{equation}
where $\underline{\bm m}^{n}, \underline{\bm w}^{n}$ denote $\underline{\bm m}(t^{n}), \underline{\bm w}(t^{n})$, which leads to the following time-stepping for the $j-$th computational cell:
\begin{align}
    \frac{{\bm m_j}^{n+1}- {\bm m_j}^n}{\Delta t}&=\bm w_j^{n+\frac{1}{2}} \times \bm m_j^{n+\frac{1}{2}} \quad, \label{eq:iLLG_m_mp}\\
    \xi\frac{\bm w_j^{n+1}-\bm w_j^n}{\Delta t}&=-\bm m_j^{n+\frac{1}{2}}\times \bm w_j^{n+\frac{1}{2}} -\alpha \bm w_j^{n+\frac{1}{2}} \nonumber \\  + \bm m_j^{n+\frac{1}{2}}\times& \bm h_\mathrm{eff,j}(\underline{\bm m}^{n+\frac{1}{2}},t^{n+\frac{1}{2}}) \,,\, j=1,\ldots,N\,. \label{eq:iLLG_w_mp}
\end{align}
Now, by dot-multiplying the first equation by $\bm m_j^{n+1/2}$, one can easily see that 
\begin{equation}
|\bm m_j^{n+1}|^2-|\bm m_j^{n}|^2=0 \,,\quad j=1,\ldots,N\,, \label{eq:IMP_ampl_cons}
\end{equation}
meaning that magnetization amplitude will be preserved unconditionally, namely independently of $\Delta t$ in each computational cell. 
In addition, by dot-multiplying Eq.~\eqref{eq:iLLG_m_mp} by $\bm w^{n+\frac{1}{2}}$ and Eq.~\eqref{eq:iLLG_w_mp} by $\bm m^{n+\frac{1}{2}}$ and summing both sides of the result, one immediately ends up with
the reproduction of the property \eqref{eq:semidiscrete exponential wdotm} in its midpoint time-discretized version:
\begin{equation}\label{eq:IMP_exponential_wdotm2}
 \frac{\bm w_j^{n+1}\cdot\bm m_j^{n+1}-\bm w_j^n\cdot \bm m_j^n}{\Delta t}= -\frac{\alpha}{\xi} \bm w_j^{n+\frac{1}{2}}\cdot \bm m_j^{n+\frac{1}{2}} \,, j=1,\ldots,N   \,.
\end{equation}
Remarkably enough, in the conservative case $\alpha=0$, the latter equation implies:
\begin{equation}\label{eq:IMP_exponential_wdotm3}
 \bm w_j^{n+1}\cdot \bm m_j^{n+1}=\bm w_j^{n}\cdot \bm m_j^{n}\,,\quad j=1,\ldots,N   
\end{equation}
which means that the 'angular momentum' projection conservation property is fulfilled for any choice of the time step $\Delta t$. 

Now let us consider the midpoint rule time-stepping for the mesh vectors:
\begin{align}
    \frac{{\underline{\bm m}}^{n+1}- {\underline{\bm m}}^n}{\Delta t}&=\underline{\Lambda}(\underline{\bm w}^{n+\frac{1}{2}}) \cdot \underline{\bm m}^{n+\frac{1}{2}} \quad, \label{eq:mesh iLLG_m_mp}\\
    \xi\frac{\underline{\bm w}^{n+1}-\underline{\bm w}^n}{\Delta t}&=-\underline{\Lambda}(\underline{\bm m}^{n+\frac{1}{2}})\cdot \underline{\bm w}^{n+\frac{1}{2}} \\
    &-\alpha \underline{\bm w}^{n+\frac{1}{2}} + \underline{\Lambda}(\underline{\bm m}^{n+\frac{1}{2}})\cdot \underline{\bm h}_\mathrm{eff}(\underline{\bm m}^{n+\frac{1}{2}},t^{n+\frac{1}{2}}) \,. \label{eq:mesh iLLG_w_mp}
\end{align}

By assuming constant applied field, dot-multiplying the second equation by $\underline{\bm w}^{n+1/2}$ and taking into account Eq.~\eqref{eq:discrete effective field} and the self-adjointness of the matrix $\underline{C}$ with respect to the adopted discrete inner product, one obtains the following discretized energy balance:
\begin{align}
    \frac{\xi}{2}\frac{|\underline{\bm w}^{n+1}|^2-|\underline{\bm w}^n|^2}{\Delta t}&= -\alpha \bm |\underline{\bm w}^{n+\frac{1}{2}}|^2 - \frac{g^{n+1}-g^n}{\Delta t} \Leftrightarrow  \\ \nonumber
    \frac{\underline{\tilde{g}}^{n+1}-\underline{\tilde{g}}^n}{\Delta t}&= -\alpha \bm |\underline{\bm w}^{n+\frac{1}{2}}|^2 \quad, \label{eq:IMP_en_bal} 
\end{align}
where $\underline{\tilde{g}}^n=\underline{\tilde{g}}(\underline{\bm m}^n,\underline{\bm w}^n)$, which implies that the total (discrete) energy $\underline{\tilde{g}}$ must be either decreasing when $\alpha>0$ or being conserved when $\alpha=0$, both regardless of the time-step.

Equations \eqref{eq:mesh iLLG_m_mp}-\eqref{eq:mesh iLLG_w_mp} represent a nonlinear system of $6N$ coupled scalar equations, which must be solved at each time step. They can be regarded as the following two vector equations in $\underline{\bm u}=\underline{\bm m}^{n+1}, \underline{\bm v}=\underline{\bm w}^{n+1}$:
\begin{equation}
    \bm F(\underline{\bm u},\underline{\bm v})=\bm 0 \quad,\quad \bm G(\underline{\bm u},\underline{\bm v}) =\bm 0 \,,
\end{equation}
and can be solved by using Newton-Raphson iteration:
\begin{align}
    \left( \begin{array}{c} \underline{\bm u}_{k+1}  \\ \underline{\bm v}_{k+1} \end{array}\right) = \left( \begin{array}{c} \underline{\bm u}_{k}  \\ \underline{\bm v}_{k} \end{array} \right) -
    \left( \begin{array}{cc} 
    \frac{\partial \bm F}{\partial \underline{\bm u}} & \frac{\partial \bm F}{\partial \underline{\bm v}} \\ 
    \frac{\partial \bm G}{\partial \underline{\bm u}} & \frac{\partial \bm G}{\partial \underline{\bm v}} \end{array} \right)^{-1} \cdot  \left( \begin{array}{c} \bm F(\underline{\bm u}_{k},\underline{\bm v}_{k})  \\ \bm G (\underline{\bm u}_{k},\underline{\bm v}_{k}) \end{array}\right) \,,
\end{align}
where the partial Jacobian matrices are given by:
\begin{align}
    \frac{\partial \bm F}{\partial \underline{\bm u}} &= \frac{\underline{I}}{\Delta t} -\frac{1}{4}\underline{\Lambda}(\underline{\bm v}+\underline{\bm w}^n) \,,\label{eq:IMR Jacobian1}\\
    \frac{\partial \bm F}{\partial \underline{\bm v}} &=  \frac{1}{4}\underline{\Lambda}(\underline{\bm u}+\underline{\bm m}^n) \,,\label{eq:IMR Jacobian2} \\
    \frac{\partial \bm G}{\partial \underline{\bm u}} &=  -\frac{1}{4}\underline{\Lambda}(\underline{\bm v}+\underline{\bm w}^n) + \frac{1}{4}\underline{\Lambda}(\underline{\bm u}+\underline{\bm m}^n)\cdot \underline{C}  \\ &+\frac{1}{2} 
    \underline{\Lambda}\left[\underline{\bm h}_\mathrm{eff}\left(\frac{\underline{\bm u}+\underline{\bm m}^n}{2}\right)\right] \,, \label{eq:IMR Jacobian3}\\
        \frac{\partial \bm G}{\partial \underline{\bm v}} &= \frac{\xi }{\Delta t}\,\underline{I} +\frac{1}{4}\underline{\Lambda}(\underline{\bm u}+\underline{\bm m}^n) + \frac{\alpha}{2}\,\underline{I} \,, \label{eq:IMR Jacobian4}
\end{align}
and the linear operator notation $\underline{\Lambda}$ has been used for the cross product involving mesh vectors.

The above time-stepping has remarkable qualitative properties that reproduce those of the continuous iLLG dynamics and, therefore, provides a structure-preserving choice for the
numerical simulation of inertial magnetization dynamics.  

However, it requires doubling the number of state variables
and, consequently, the number of unknowns, owing to the introduction of the vector field $\bm w$ although one is mainly interested in computing the dynamics of
the magnetization vector field $\bm m$. This issue becomes even more pronounced when large-scale micromagnetic simulations with full spatial discretization are considered, which would require substantial modifications to existing
numerical codes in order to introduce the auxiliary variable $\bm w$ and to solve a system of $6N$ nonlinear coupled equations at each time-step. Moreover, we remark that in the latter situation, the $3N\times 3N$ matrix $\underline{C}$ involved in the Newton iteration (see Eq.~\eqref{eq:IMR Jacobian3}) is also fully-populated owing to the long-range nature of magnetostatic interactions. Therefore, following what is done for the classical LLG equation\cite{dAquino2005}, a quasi-Newton technique is required to solve the large nonlinear system, implemented by considering reasonable sparse approximation of the matrix $\underline{C}$ (e.g. obtained retaining only exchange and anisotropy contributions $\underline{C}_\mathrm{ex}$ and $\underline{C}_\mathrm{an}$, respectively) and, in turn, of the full Jacobian defined by Eqs.~\eqref{eq:IMR Jacobian1},\eqref{eq:IMR Jacobian2},\eqref{eq:IMR Jacobian3},\eqref{eq:IMR Jacobian4}. Of course, the computational cost of such quasi-Newton method, involving the solution of several non-symmetric large linear systems (e.g. by using GMRES methods\cite{Saad1986}), will be at least twice that associated with the
classical LLG equation, posing a strong limit to the capability of solving large-scale iLLG dynamics.


For these reasons, an alternative efficient numerical
technique requiring only minor modifications to existing
micromagnetic codes was proposed in
Ref.~\cite{daquino2024midpoint}. This second time-stepping is based on the implicit
midpoint rule combined with a multi-step method
(hereafter referred to as IMR-MS) for the inertial term. This technique is based on direct space-time discretization of Eq.~\eqref{eq:iLLG equation} at time $t^{n+\frac{1}{2}}=t_n+\Delta t/2$:
\begin{align}
    \frac{\underline{\bm m}^{n+1}- \underline{\bm m}^n}{\Delta t}&= - \underline{\Lambda}(\underline{\bm m}^{n+\frac{1}{2}})\cdot \left(\underline{\bm h}_\mathrm{eff}\left(\underline{\bm m}^{n+\frac{1}{2}},t^{n+\frac{1}{2}}\right) \right. \\&-\alpha \frac{\underline{\bm m}^{n+1}- \underline{\bm m}^n}{\Delta t}
    -\xi \frac{d^2\underline{\bm m}}{dt^2}\left|_{n+\frac{1}{2}} \right) \,. \label{eq:iLLG_mp}
\end{align}

Then, in order to retain the amplitude conservation property \eqref{eq:IMP_ampl_cons} of the aforementioned IMR scheme, we use the first of midpoint formulas \eqref{eq:mp formulas} in Eq.~\eqref{eq:iLLG_mp} in a way that it is rewritten as system of $3N$ nonlinear equations in the unknowns $\underline{\bm m}^{n+1}$:
\begin{equation}\label{eq:nonlinear_system}
\bm F^n(\underline{\bm m}^{n+1})=\mathbf{0}\quad,
\end{equation}
where $\bm F^n(\underline{\bm y}):\mathbb{R}^{3N}\rightarrow \mathbb{R}^{3N}$
is the following vector function:
\begin{equation}\label{eq:nonlinear_function}
\begin{split}
\bm F^n(\underline{\bm y})&=\left[I-\alpha\underline{\Lambda}\left(\frac{\underline{\bm y}+\underline{\bm m}^{n}}{2}\right)\right]
\left(\underline{\bm y}-\underline{\bm m}^{n}\right) - \Delta t
\,\,\\ &\bm f^n\left(\frac{\underline{\bm y}+\underline{\bm m}^{n}}{2}\right) -\Delta t\, \xi \underline{\Lambda}\left(\frac{\underline{\bm y}+\underline{\bm m}^{n}}{2}\right)\frac{d^2\underline{\bm m}}{dt^2}\left|_{n+\frac{1}{2}} \right. \,,
\end{split}
\end{equation}
and where
\begin{equation}\label{eq:f_conservative_flow}
   \bm f^n(\underline {\bm m})=-\underline{\Lambda}(\underline{\bm m}) \cdot \underline{\bm h}_\text{eff}\left(\underline{\bm m},t^n+\frac{\Delta t}{2}\right)
\end{equation}
is the purely precessional term in the right-hand side obtained from the direct semi-discretization of the iLLG Eq.~\eqref{eq:iLLG equation}, where the discrete effective field is expressed by using the definition \eqref{eq:discrete effective field}.

The IMR-MS method is based on the adoption of a multi-step approach with a $p-$point backward formula for the second derivative in the inertial term that appears in Eqs.~\eqref{eq:iLLG_mp} and \eqref{eq:nonlinear_function}:
\begin{equation}\label{eq:d2m multi step}
    \frac{d^2\underline{\bm m}}{dt^2}\left|_{n+\frac{1}{2}} \right. \approx
\frac{1}{\Delta t^2}\sum_{k=1}^{p} a_{n+2-k} \underline{\bm m}^{n+2-k} =\Delta^2_{p}\,,\,p\geq3
\end{equation}
where the coefficients $a_{n+2-k}$ are determined from truncation error analysis in order to control the accuracy of the approximation.
This technique implies a slight modification of existing numerical codes based on implicit midpoint rule time-stepping. In fact, once formula \eqref{eq:d2m multi step} is plugged into the time-stepping equation \eqref{eq:iLLG_mp}, the solution of the nonlinear coupled equations \eqref{eq:nonlinear_system} can be obtained by using the Newton-Raphson technique\cite{dAquino2005} as follows:
\begin{equation}
\begin{split}
&\underline{\bm y}_0=\underline{\bm m}^n  \,  , \,\,\,\,
\underline{\bm y}_{k+1}=\underline{\bm y}_k+\Delta\underline{\bm y}_{k+1} \, \\
&J^n_F(\underline{\bm y}_k,t^n)\Delta\underline{\bm y}_{k+1}=-\bm F^n(\underline{\bm y}_k) \,, \label{eq:iter}
\end{split}
\end{equation}
by simply considering the following augmented Jacobian matrix of the iteration:
\begin{align}
    J_F(\underline{\bm u},t)&=\frac{\underline{I}}{\Delta t}-\frac{\alpha}{\Delta t}\underline{\Lambda}(\underline{\bm m}^n) \\
    &+\frac{\xi}{2\Delta t^2}a_1\underline{\Lambda}(\underline{\bm u}) +\frac{\xi}{2\Delta t^2}\underline{\Lambda}\left(\sum_{k=2}^{p} a_{n+2-k} \underline{\bm m}^{n+2-k}\right) \nonumber \\&-\frac{\xi}{2\Delta t^2}\underline{\Lambda}(\underline{\bm u} + \underline{\bm m}^n) a_1-\frac{1}{2}J_f\left(\frac{\underline{\bm u}+ \underline{\bm m}^n}{2},t+\frac{\Delta t}{2}\right) \label{eq:IMR timestepping Jacobian}
\end{align}
where $J_f(\underline{\bm u},t)=\underline{\Lambda}(\underline{\bm u})\cdot \underline{C}+\underline{\Lambda}[\underline{\bm h}_\mathrm{eff}(\underline{\bm u},t)]$.
The linear system in Eq.~\eqref{eq:iter} is solved at each iteration $k$ by considering the sparse approximation of the full matrix $\underline{C}$ as $\underline{C}\approx \underline{C}_\mathrm{ex}+\underline{C}_\mathrm{an}$ in the Jacobian $J_f$ and using the GMRES method\cite{Saad1986} until the residual $\|\bm F^n(\underline{\bm y}_k)\|$ decreases below a prescribed tolerance. 

\begin{table}[t]
    \centering
    \begin{tabular}{|c|c|c|c|}
    \hline 
    scheme & Coefficients  \\
    \hline
    IMR-MS1 & $a_1=1, a_0=-2, a_{-1}=1$  \\
    IMR-MS2 & $a_1=3/2, a_0=-7/2, a_{-1}=5/2, a_{-2}=-1/2$ \\
     \hline
    \end{tabular}
    \caption{Table of coefficients for multi-step formula \eqref{eq:d2m multi step}.}
    \label{tab:tab1}
\end{table}

Now, if one assumes that initially magnetization has zero velocity $d\underline{\bm m}/dt(t=0)=\bm 0$, which is reasonable in simulation of experimental situations, then at the first time step $n=1$ one has $\bm m^{n+2-k}=\bm m^0, k>2$. For the subsequent steps $n>1$, one can use magnetization samples from the previous steps as $\underline{\bm m}^{n+2-k}, k>2$ in Eq.~\eqref{eq:iLLG_mp}. The only additional cost of this procedure is the
storage of $p-2$ magnetization vectors.

The technical analysis performed in Ref.~\cite{daquino2024midpoint} leads to the following two formulas, which are first and second-order accurate, respectively:
\begin{align} 
    \frac{d^2\underline{\bm m}}{dt^2}\left|_{n+\frac{1}{2}} \right. &\approx \frac{\underline{\bm m}^{n+1} + \underline{\bm m}^{n-1}- 2 \underline{\bm m}^n}{\Delta t^2} = \Delta^2_{p=3}\,, \label{eq:D2p3} \\
    \frac{d^2\underline{\bm m}}{dt^2}\left|_{n+\frac{1}{2}} \right. &\approx \frac{3\underline{\bm m}^{n+1}-7\underline{\bm m}^n+5 \underline{\bm m}^{n-1}-\underline{\bm m}^{n-2}}{2\Delta t^2} = \Delta^2_{p=4}\,, \label{eq:D2p4}
\end{align}
which, plugged into Eq.~\eqref{eq:iLLG_mp}, define the IMR-MS1 and IMR-MS2 schemes, respectively.

The coefficients for the above multi-step formulas with $p=3,4$ are reported in Table~\ref{tab:tab1} and can be straightforwardly used in Eqs.~\eqref{eq:nonlinear_function} and \eqref{eq:IMR timestepping Jacobian}.

\subsection{Examples of time-domain micromagnetic simulations with inertia}

In this section, we report the results of simulations of transient inertial magnetization dynamics in confined thin-film ferromagnetic nanodots. In order to perform efficient time-domain micromagnetic simulations of iLLG dynamics with full spatial discretization, the IMR-MS2 time-stepping has been implemented in the finite-difference numerical code MaGICo\cite{dAquino2005,MaGICo}, with a computational cost comparable to that of classical precessional LLG dynamics while keeping important conservation properties as outlined in the previous sections. The first numerical experiment complements the frequency-domain study reported in section \ref{sec:frequency domain mumag} by showing finite delay effects in the propagation of inertial spin waves\cite{daquino2023micromagnetic} driven by transient ac terahertz field pulses. The second example addresses the strongly nonlinear dynamics pertaining to ultrafast magnetization switching in the inertial regime\cite{neeraj2022inertial}, for which very recently an analytical solution has been proposed\cite{Fortunati_2026} within the macrospin approximation.  

\subsubsection{Spatially-inhomogeneous spin nutation driven by terahertz applied field}

\begin{figure*}[t!]
    \centering
    \includegraphics[width=15cm]{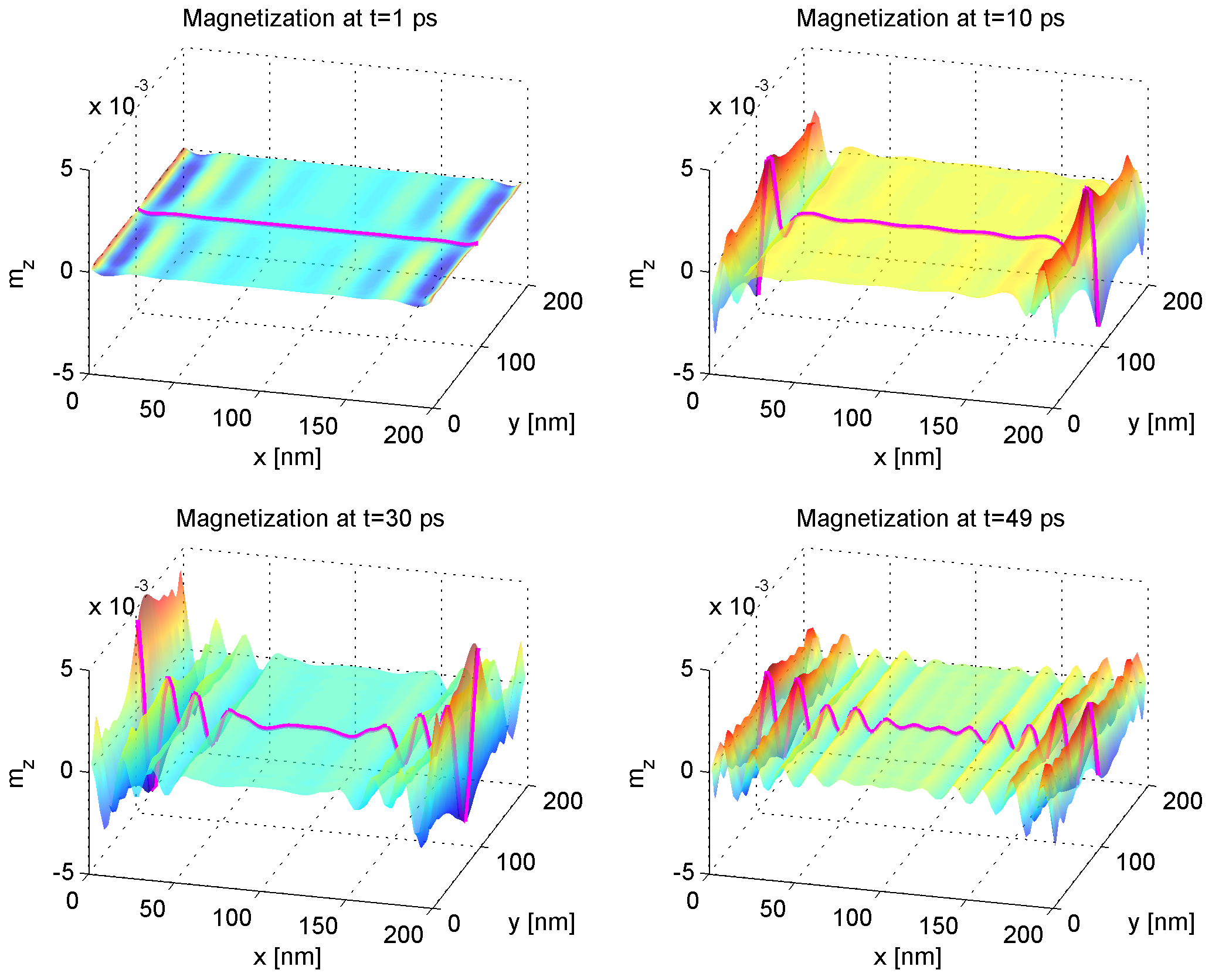}
    \caption{Spatial profiles of short-wavelength magnetization out-of-plane component $m_z$ obtained by FFT high-pass filtering at times $t=1,10,30,49$~ps. The magenta solid line is a guide for the eye to follow the spin wave oscillation profile along the $x$ direction. (Reproduced from Ref.~\cite{daquino2024midpoint}
under the terms of the Creative Commons Attribution
4.0 licence.)}
    \label{fig:spin waves}
\end{figure*}

Here we explore typical spatio-temporal patterns of iLLG dynamics considering the time-domain simulation of ultrashort inertial spin waves in a confined ferromagnetic thin film. At terahertz frequencies, the behavior of small magnetization oscillations significantly deviates from the classical description of exchange-dominated spin waves, in that an ultimate limiting propagation speed appears\cite{daquino2023micromagnetic}. This difference is mostly due to the mathematical structure of Eq.~\eqref{eq:iLLG equation} compared with the classical LLG precessional dynamics, i.e. the same equation where one sets $\xi=0$. In fact, when inertial effects are taken into account, the torque proportional to the second-order time-derivative transforms the classical LLG equation into a wave-like equation with hyperbolic mathematical character. In this respect, on short time scales, finite time delays are expected in magnetization response propagation away from the region of local external excitation. 

The considered sample is made of cobalt and has a thin-film shape with square cross-section $200\times 200\,$nm$^2$ and thickness $5\,$nm. The ferromagnetic nanodot is initially at equilibrium, being saturated along the $x$ axis by a static field $\mu_0 H_{ax}=100$ mT. 
The values of the material parameters are $\gamma=2.211\times 10^5$ m$\,$A$^{-1}\,$s$^{-1}$, $\mu_0 M_s=1.6$ T, $A=13$ pJ/m ($l_\mathrm{ex}=3.57$ nm), $t_\mathrm{in}=0.653$ ps $(\xi=0.0338)$ and $\alpha=0.005$.

The applied field is a spatially-uniform sine wave step (turned on at $t=0$) along the $y$ axis transverse to the equilibrium configuration with amplitude $\mu_0 H_{ay}=100$ mT and frequency $f=1386$ GHz. For the above choice of parameters, this excitation frequency is slightly above the nutation resonance frequency and corresponds to inertial spin waves with wavelength around 20 nanometers\cite{daquino2023micromagnetic}.  

The numerical simulation of iLLG equation \eqref{eq:iLLG equation} is performed with a time-step of 25 fs, a $2.5\times 2.5\times 5$ nm$^3$ computational cell, which corresponds to discretizing the thin film into $80\times 80$ square-prism cells. To isolate short-wavelength spin-wave propagation from the remaining spatial response, high-pass
filtering based on a two-dimensional Fast Fourier Transform is applied to the magnetization components. The simulated time is $100\,$ps and some snapshots of the magnetization out-of-plane component $m_z$, taken at different times, are reported in Fig.~\ref{fig:spin waves}. 

The magnetization is initially at equilibrium and mostly aligned with the static field along the $x$ direction except close to the square corners where there is the most pronounced deviation. Such a tilting acts as a local excitation for inertial spin waves when the time-varying ac field step is applied\cite{daquino2023micromagnetic}. In fact, as shown in the various panels of Fig.~\ref{fig:spin waves}, two wave packets with wavelength $\approx 20\,$nm propagate from the edges of the nanodot toward its center at a finite speed of
approximately $2000$~m/s after the application of the ac field, consistently with the theoretical prediction\cite{daquino2023micromagnetic}.

\subsubsection{Micromagnetic simulation of inertial switching}

We now consider the second numerical example anticipated above,
namely the strongly nonlinear magnetization switching process in
the inertial regime investigated in
Ref.~\cite{neeraj2022inertial}. As recently discussed in
Ref.~\cite{Fortunati_2026}, this process admits an
approximate analytical description within the macrospin
approximation. Here, micromagnetic simulations are used to assess
the validity of such a spatially coherent description when the
macrospin assumption is relaxed and spatial inhomogeneities are
fully taken into account.

To this end, we consider a permalloy disk-shaped nanodot with diameter $D=1\,\mu m$  and thickness of $d=5$ nm. The very small aspect ratio $d/D=1/200$ of such a nanodot allows one to approximate the infinite thin-film limit used in the previous macrospin calculation, yielding effective demagnetizing factors $N_1=N_2\approx 10^{-2}, N_3\approx 0.98$ that are close enough to the ideal ones $N_1=N_2=0,N_3=1$. The material parameters are the same as those used in Ref.~\cite{neeraj2022inertial}: $\mu_0 M_s=0.92$~T, $\xi=0.02$, $\alpha=0.01$, exchange stiffness constant $A_\mathrm{ex}=13$ pJ/m (exchange length $\ell_\mathrm{ex}=6.21$ nm), uniaxial anisotropy along the $x$ direction corresponding
to an effective anisotropy field of $100$~mT. The disk is discretized on a $200\times 200$ regular grid into a collection of 31428 computational cells with dimension $5\times5\times5$~nm$^3$. The inertial micromagnetic simulations are performed using the large-scale midpoint geometric integrator described in Ref.~\cite{daquino2024midpoint} and implemented in the micromagnetic solver MaGICo\cite{MaGICo}.

The initial equilibrium configuration of the disk is obtained by relaxing the magnetization under zero external field from an initial state $m_1(t=0)=1$, using a (fictitious) large Gilbert damping $\alpha=1$ until the (dimensionless) magnetic effective torque becomes smaller than $10^{-5}$. It turns out that, due to shape and crystalline magnetic anisotropy, the equilibrium configuration is single-domain with magnetization mostly aligned with the $x$ axis, yielding $\langle m_1 \rangle=0.9995$ (the notation $\langle\cdot\rangle$ denotes spatial average over the disk volume).

\begin{figure*}[t!]\begin{center}
(a) \hspace{0.5\textwidth} (b) \\
\includegraphics[width=0.45\textwidth]{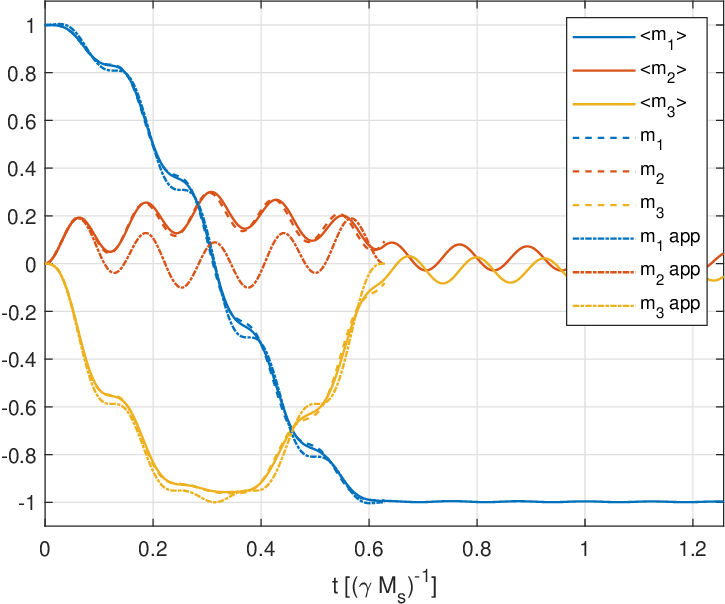} \hfill\includegraphics[width=0.45\textwidth]{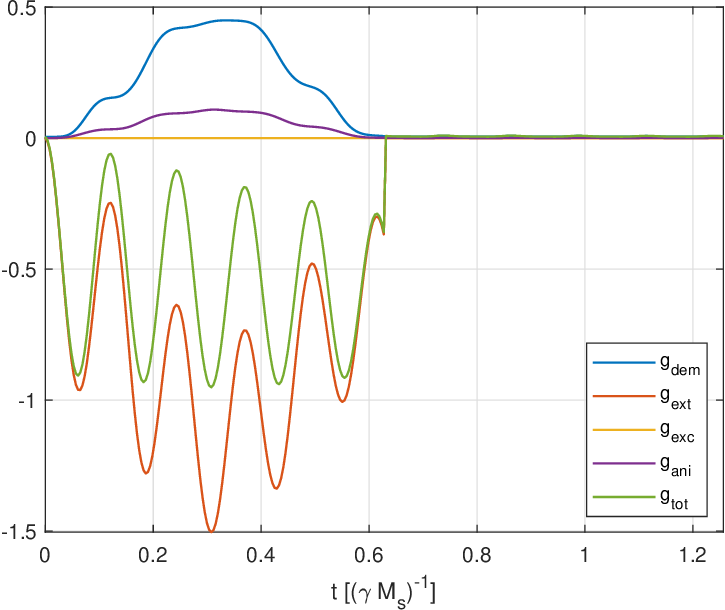}
  \end{center}
	\caption{Micromagnetic simulation of inertial switching. In panel (a), the evolution of the components of the spatially-averaged magnetization $\langle m_1\rangle, \langle m_2\rangle,\langle m_3\rangle$ (solid lines) for $t \in [0, 2 T_{sw}]$ are compared with those obtained from macrospin (dashed) and approximate (dashed-dotted) analytical solution developed in Ref.~\cite{Fortunati_2026}, where $T_{sw}=0.6283$ is the switch-off instant of the applied field. The analytical solution is only plotted for $t\in[0,T_{sw}]$. Panel (b) shows the micromagnetic free energy dynamics, separated into magnetostatic $g_\mathrm{dem}$, Zeeman $g_\mathrm{ext}$, exchange $g_\mathrm{exc}$, anisotropy $g_\mathrm{ani}$ and total $g_\mathrm{tot}$ in dimensionless units of $\mu_0 M_s^2 V$ ($V=d \pi D^2/4$ is the disk volume). (Reproduced from Ref.~\cite{Fortunati_2026}
under the terms of the Creative Commons Attribution
4.0 licence.)}
	\label{fig:validation4}
	\vspace{0pt}
\end{figure*}

The switching simulation is started from the above
equilibrium state by applying a rectangular external-field
pulse along the in-plane transverse direction with amplitude $\mu_0 H_a=4.6~$T ($h_a=5$ in dimensionless units) and duration equal to $(\gamma M_s)^{-1}T_{sw}=3.88$ ps. The time-step used in the simulation is 25 fs, which ensures accurate resolution of the ultrafast nutation dynamics. The results are reported in Fig.~\ref{fig:validation4}. In panel (a), the time-evolution of the spatially averaged magnetization components $\langle m_1\rangle, \langle m_2\rangle,\langle m_3\rangle$ is plotted (solid lines) for an interval equal to twice the field pulse duration $2T_{sw}$. This dynamics can be directly compared with the one computed under the macrospin assumption as well as with the approximate analytical solution (dashed and dashed-dotted lines in Fig.~\ref{fig:validation4}(a), respectively), which exhibits excellent agreement especially concerning the switching variable $m_1(t)$. 

This result can be understood by examining the
free-energy evolution shown in Fig.~\ref{fig:validation4}(b), where one can see that the exchange energy associated with spatial inhomogeneities (yellow line) remains close to zero during the whole switching process. Conversely, one can see that, during the switching driven by the external field pulse, the dominant energy terms are the Zeeman and magnetostatic ones followed by the uniaxial anisotropy with each being much larger than the exchange contribution. Thus, one can conclude that, in the analyzed situation, the precessional magnetization switching in the inertial regime essentially occurs through spatially coherent dynamics.

\section{Conclusions and perspectives}
\label{sec:conclusions}

Magnetization dynamics at sub-picosecond time scales can enter a
regime in which magnetic inertia supplements the conventional
Landau--Lifshitz--Gilbert description. The resulting inertial
Landau--Lifshitz--Gilbert (iLLG) equation introduces an additional
characteristic time scale associated with angular-momentum relaxation
and gives rise to magnetization nutation at THz frequencies,
coexisting with the conventional, much slower precessional dynamics.
Following its theoretical prediction and experimental observation,
magnetic inertia has therefore emerged as an important ingredient in
the description of ultrafast magnetization dynamics.

In this topical review, we have discussed the modelling and numerical
simulation of inertial magnetization dynamics with particular emphasis
on spatially inhomogeneous ferromagnetic systems. The mechanical
analogy underlying the iLLG equation provides a useful physical
interpretation of magnetic inertia, while its formulation as a
first-order dynamical system in terms of magnetization and an
additional angular-momentum variable makes transparent its main
qualitative properties. These include conservation of the
magnetization amplitude, angular-momentum relaxation and a generalized
free-energy balance in which the usual micromagnetic free energy is
supplemented by an inertia-related contribution playing the role of a
kinetic energy.

The linearized iLLG equation provides a unified framework for
understanding the spectral properties of inertial magnetization
dynamics. Magnetic inertia modifies the conventional precessional
spin-wave spectrum and introduces an additional nutational branch
typically located in the THz range. In the weak-inertia regime, the
precessional modes undergo an inertial redshift, while the spectral
linewidths of both branches are also affected by inertia. These
features have been discussed for macrospin systems and for the
classical backward-volume, Damon--Eshbach and forward-volume
spin-wave geometries in extended magnetic thin films. In curved and
twisted magnetic nanostructures, geometrical curvature and torsion
provide additional degrees of freedom, with torsion giving rise to
nonreciprocal propagation and opposite dispersion asymmetries for
precessional and nutational modes.

Spatially resolved modelling becomes particularly important for the
short-wavelength regime. Exchange and magnetostatic interactions allow
nutational excitations with nanometer wavelengths to be sustained at
THz frequencies, while the hyperbolic character introduced by magnetic
inertia qualitatively modifies the large-wavenumber spin-wave
dispersion. In particular, the group velocity approaches a finite
limiting value determined by the ratio between the exchange length and
the inertial time scale, in marked contrast with the classical
exchange-dominated LLG dynamics.

Micromagnetic simulation provides the natural framework for extending
these analytical results to realistic nanostructures. Frequency-domain
methods efficiently describe linear THz resonances and inertial
spin-wave excitation, whereas time-domain simulations allow transient
propagation and strongly nonlinear processes such as ultrafast
magnetization switching to be investigated. In this context,
structure-preserving numerical methods are particularly attractive.
The geometric midpoint techniques reviewed here reproduce fundamental
constraints and energetic properties of the continuous iLLG dynamics
at the discrete level, while the midpoint--multistep formulation
provides an efficient route for incorporating magnetic inertia into
existing large-scale micromagnetic solvers. Together with complementary
finite-difference and finite-element approaches, these developments
provide an increasingly mature computational framework for inertial
micromagnetics.

The examples discussed in this review also illustrate the
complementary roles of analytical modelling and numerical simulation.
Spatially resolved calculations confirm the excitation and finite-speed
propagation of ultrashort nutational spin waves in confined
ferromagnetic nanostructures. Likewise, micromagnetic simulations of
inertial magnetization switching make it possible to assess the
validity of macrospin and analytical approximations and to establish
when the switching process remains essentially spatially coherent.

Despite this progress, several questions remain open. From the
fundamental viewpoint, establishing the magnitude, tensorial character
and microscopic origin of magnetic inertia in different materials
remains central to connecting phenomenological iLLG models with
electronic and atomistic descriptions. On the experimental side,
further measurements of nutational resonances, linewidths and
propagating inertial spin waves would provide important tests of
theoretical predictions and help determine inertial parameters with
greater accuracy. The interplay between magnetic inertia and
anisotropy, chirality, noncollinear magnetic states and geometrical
confinement also offers a broad landscape for future investigations.
At the same time, the extremely short wavelengths and high frequencies
of nutational excitations pose the practical challenge of developing
efficient mechanisms for their generation and detection.

Another intriguing direction concerns the possible use of
magnetoelastic coupling for the excitation of inertial spin waves.
Surface acoustic waves are known to efficiently excite conventional
spin waves under suitable energy- and momentum-matching conditions,
while their characteristic propagation velocities are comparable to
the limiting group velocities predicted for inertial spin waves.
Whether this mechanism can be extended to the nutational regime
through coherent acoustic excitations reaching THz frequencies and
nanometer wavelengths remains an interesting open question.

From a computational perspective, further developments will be needed
to extend inertial micromagnetics toward increasingly complex
materials, geometries and multiphysics problems while maintaining
accuracy and computational efficiency on the relevant femtosecond and
nanometer scales. In particular, the coupling of inertial
magnetization dynamics with spin transport, elastic degrees of freedom
and other ultrafast excitation mechanisms represents a natural
direction for future modelling efforts.

Overall, magnetic inertia extends micromagnetics into a dynamical
regime where GHz precession and THz nutation coexist and interact.
The combination of analytical theory, structure-preserving numerical
methods and spatially resolved micromagnetic simulations provides a
natural framework for investigating this regime and for connecting
the fundamental physics of angular-momentum relaxation with future
experiments on ultrafast magnetization dynamics and inertial
magnonics.

\bibliography{paper.bib}

\appendix

\onecolumngrid
\section{Analytical solution of the unforced linear iLLG dynamics}\label{sec:appendix}

Here we analyze the nontrivial solution of the iLLG equation written in the following general form using Fourier representation:
\begin{equation}\label{eq:general linear iLLG}
    \delta \hat{\bm m}= \left\{\mathcal{P}_\perp \left[i\omega\Lambda(\bm m_0)+\overleftrightarrow{D}(\bm k)+(h_0-\xi\omega^2+i\omega\alpha)\overleftrightarrow{I}\right]\right\}^{-1} \!\!\delta\hat{ \bm h}_\mathrm{ac} \,\,,
\end{equation}
where $\mathcal{P}_\perp=\mathcal{I}-\bm m_0\otimes\bm m_0$ is the projection operator on the plane transverse to the equilibrium $\bm m_0$, $\overleftrightarrow{D}(\bm k)=D_{pq}(\bm k)\,,\,p,q=1,2,3$ is the  $3\times 3$ wavenumber-dependent effective field tensor (including terms accounting for exchange, magnetostatics, uniaxial anisotropy) acting on magnetization perturbation in plane-wave-like form, $h_0$ is the component of the static internal field along $\bm m_0$. 

Nontrivial solutions of the unforced Eq.~\eqref{eq:general linear iLLG} are obtained by setting the determinant to zero:
\begin{equation}
\det[\overleftrightarrow{\chi}(\bm k,\omega)^{-1}]=\det\left\{\mathcal{P}_\perp \left[i\omega\Lambda(\bm m_0)+\overleftrightarrow{D}(\bm k)+(h_0-\xi\omega^2+i\omega\alpha)\overleftrightarrow{I}\right] \right\}=0 \,, \label{eq:det inv_Chi zero} 
\end{equation}
where the susceptibility tensor $\overleftrightarrow{\chi}(\bm k,\omega)$ is given by the inverse of the quantity within curly braces in Eq.~\eqref{eq:general linear iLLG}.

In order to obtain a general but simple analytical solution of Eq.~\eqref{eq:general linear iLLG}, we assume that $\bm m_0$ lies along one of the coordinate axes of the thin film. Consequently, one has that
$\mathcal{P}_\perp \Lambda(\bm m_0)=J$ becomes the $2\times2$ anti-symmetric tensor
\begin{equation}
\mathcal{P}_\perp \Lambda(\bm m_0)=
J =
\begin{pmatrix}
0 & -1 \\
1 & 0
\end{pmatrix} \,.
\end{equation}
and $\overleftrightarrow{\chi}(\bm k,\omega)^{-1}$ becomes the $2\times 2$ matrix:
\begin{equation}
\overleftrightarrow{\chi}(\bm k,\omega)^{-1} =
\begin{pmatrix}
D_{11}(\bm k)+h_0-\xi\omega^2+i\omega\alpha & D_{12}(\bm k)-i\omega \\
D_{21}(\bm k)+i\omega & D_{22}(\bm k)+h_0-\xi\omega^2+i\omega\alpha
\end{pmatrix} \,,
\end{equation}
where the subscripts $1,2$ refer to the coordinate axes in the plane transverse to $\bm m_0$. It is now convenient defining the quantities:
\begin{align}
    \omega_{11}(\bm k)=D_{11}(\bm k)+h_0 \,,\\
    \omega_{22}(\bm k)=D_{22}(\bm k)+h_0 \,,
\end{align}
which substituted in the latter equation give:
\begin{equation}
\overleftrightarrow{\chi}(\bm k,\omega)^{-1} =
\begin{pmatrix}
\omega_{11}(\bm k)-\xi\omega^2+i\omega\alpha & D_{12}(\bm k)-i\omega \\
D_{21}(\bm k)+i\omega & \omega_{22}(\bm k)-\xi\omega^2+i\omega\alpha
\end{pmatrix} \,.
\end{equation}
According to Eq.~\eqref{eq:det inv_Chi zero}, the (complex) resonance frequencies are given by the solution of the following fourth-degree polynomial:
\begin{equation}
    (\omega_{11}(\bm k)-\xi\omega^2+i\omega\alpha)(\omega_{22}(\bm k)-\xi\omega^2+i\omega\alpha)-(D_{21}(\bm k)+i\omega)(D_{12}(\bm k)-i\omega) = 0 \,,
\end{equation}
that can be rewritten as:
\begin{equation}
 \xi^2\omega^4-2i\alpha\xi\,\omega^3
-\left(1+\alpha^2+\xi\bigl(\omega_{11}+\omega_{22}\bigr)\right)\omega^2
+i\left[\alpha\bigl(\omega_{11}+\omega_{22}\bigr)+D_{21}-D_{12}\right]\omega
+\omega_{11}\omega_{22}-D_{12}D_{21} = 0 \,, \label{eq:gen characteristic poly}
\end{equation}
where the dependence on $\bm k$ is understood. For the sake of simplicity, we assume here that the effective field tensor is diagonal, namely $D_{12}=D_{21}=0$. Under this assumption, Eq.~\eqref{eq:gen characteristic poly} becomes:
\begin{equation}
 \xi^2\omega^4-2i\alpha\xi\,\omega^3
-\left(1+\alpha^2+\xi\bigl(\omega_{11}+\omega_{22}\bigr)\right)\omega^2
+i\left[\alpha\bigl(\omega_{11}+\omega_{22}\bigr)\right]\omega
+\omega_{11}\omega_{22} = 0 \,. \label{eq:gen characteristic poly sym}
\end{equation}
A perturbation analysis provides the following 4 roots $\pm \omega_N+i\Delta\omega_N$ and $\pm\omega_P+i\Delta\omega_P$:
\begin{align}
  \omega_N&
  \approx\sqrt{\frac{1+\xi(\omega_{11}+\omega_{22}) +\sqrt{1+2\xi(\omega_{11}+\omega_{22}) }}{2\,\xi ^2}} \label{eq:gen pert omega_N} \\
    \omega_P&\approx 
\frac{\sqrt{\omega_{11}\,\omega_{22}}}{\sqrt{1+\xi(\omega_{11}+\omega_{22}) }} \label{eq:gen pert omega_P} \\
    \Delta\omega_N&\approx\left(1+\frac{1}{\sqrt{1+2\xi(\omega_{11}+\omega_{22}) }} \right)\,\frac{\alpha}{2\xi} \,,  \label{eq:gent pert delta_omega_N} \\
    \Delta\omega_P&\approx\left(1-\frac{1}{\sqrt{1+2\xi(\omega_{11}+\omega_{22}) }}\right)\,\frac{\alpha}{2\xi} \,. \label{eq:gent pert delta_omega_P}\\    
\end{align}
In deriving Eqs. \eqref{eq:gen pert omega_N}-\eqref{eq:gen pert omega_P} and the corresponding linewidths \eqref{eq:gent pert delta_omega_N}-\eqref{eq:gent pert delta_omega_P}, terms of order $\mathcal{O}(\xi^2)$ in the conservative characteristic quantities are neglected before performing the first-order expansion in damping.
The last expressions can be conveniently simplified if $\xi\ll 1$ and moderately small wavenumber $|\bm k|$:
\begin{align}
   \omega_N&\approx \frac{1}{\xi} + \frac{1}{2}(\omega_{11}+\omega_{22}) \,, \label{eq:gen pert omega_N small xi} \\
    \omega_P&\approx \sqrt{\omega_{11}\omega_{22}} \left[1- \frac{\xi}{2}(\omega_{11}+\omega_{22})\right] \,, \label{eq:gen pert omega_P small xi} \\
   \Delta\omega_N
&\approx
\frac{\alpha}{\xi}
-\frac{\alpha}{2}(\omega_{11}+\omega_{22})
+\frac{3}{4}\alpha\xi(\omega_{11}+\omega_{22})^2 \,, \label{eq:gen pert delta_omega_N small xi}
\\ 
    \Delta\omega_P
&\approx
\frac{\alpha}{2}(\omega_{11}+\omega_{22})
-\frac{3}{4}\alpha\xi(\omega_{11}+\omega_{22})^2= \frac{\alpha}{2}(\omega_{11}+\omega_{22})\left[1-\frac32\xi(\omega_{11}+\omega_{22})\right]\,. \label{eq:gen pert delta_omega_P small xi}
\end{align}


\end{document}